%% file: NonStab.tex
\documentclass[12pt]{article}
\usepackage{graphics}
\usepackage{fullpage}
\usepackage{amssymb}
\usepackage{amsfonts}
\usepackage{amsmath}
\usepackage[mathscr]{eucal}

\newtheorem{theorem}{Theorem}[section]
\newtheorem{lemma}[theorem]{Lemma}

\newtheorem{corollary}[theorem]{Corollary}
\newtheorem{claim}[theorem]{Claim}
\newtheorem{proposition}[theorem]{Proposition}
\newtheorem{observation}[theorem]{Observation}
\newtheorem{definition}[theorem]{Definition}
\newtheorem{problem}[theorem]{Problem}
\newtheorem{remark}[theorem]{Remark}

\newcommand{\color}[2][]{}

\newcommand
{\floor}[1]{\ensuremath{\left\lfloor{#1}\right\rfloor}}
\newcommand {\ket} [1] {\ensuremath \left\vert#1\ensuremath \right\rangle}
\newcommand {\bra} [1] {\ensuremath \left\langle#1\ensuremath \right\vert}
\newcommand {\braket} [2]
{\ensuremath \left\langle#1\ensuremath\,{\vert}\,{#2}\ensuremath \right\rangle}

\newcommand{\F}{\ensuremath{\mathbb{F}}}
\newcommand{\GF}[2][]{{\ensuremath{\mathbb{F}_{#2}^{#1}}}}

\newcommand{\GL}[2][2]{\ensuremath{GL_{#2}(\mathbb{F}_{#1})}}
\newcommand{\GLnq}{\GL[q]{m}}

\newcommand{\bproof}{\noindent{\it Proof}}
\newcommand{\cproof}{\noindent{\it Proof of Claim}}
\newcommand{\eproof}{\hspace*{\fill}$\rule{2mm}{2mm}$~~~~~\bigskip}
\renewenvironment{proof}{\bproof. }{\eproof}

\newenvironment{claimproof}{\cproof. }{\hspace*{\fill}\vspace{5mm}}

\newcommand{\Tr}{\mbox{\it Tr}}

\newcommand{\Ints}{{\mathbb{Z}}}

\renewcommand{\bold}[1]{{\bf #1}}

\renewcommand{\a}{{\bf a}}
\renewcommand{\b}{{\bf b}}
\renewcommand{\c}{{\bf c}}
\renewcommand{\d}{{\bf d}}
\newcommand{\e}{{\bf e}}
\newcommand{\x}{{\bf x}}
\renewcommand{\v}{{\bf v}}
\renewcommand{\u}{{\bf u}}
\newcommand{\y}{{\bf y}}

\newcommand{\om}{{\omega}}
\newcommand{\w}{\tilde{\omega}}

\newcommand{\balpha}{\ensuremath{\mbox{\boldmath $\alpha$}}}
\newcommand{\bbeta}{\ensuremath{\mbox{\boldmath $\beta$}}}
\newcommand{\bchi}{\ensuremath{\mbox{\boldmath $\chi$}}}
\newcommand{\sbalpha}{\ensuremath{\mbox{\scriptsize\boldmath $\alpha$}}}
\newcommand{\sbbeta}{\ensuremath{\mbox{\scriptsize\boldmath $\beta$}}}
\newcommand{\sbchi}{\ensuremath{\mbox{\scriptsize\boldmath $\chi$}}}

\newcommand{\brac}[2]{{\ensuremath{[[{#1}]]_{{#2}}}}}

\newcommand{\wt}{{\it wt}}

\newcommand{\Hi}{{\ensuremath{\mathcal{H}}}}
\newcommand{\A}{{\ensuremath{\mathcal{A}}}}
\newcommand{\B}{{\ensuremath{\mathcal{B}}}}
\newcommand{\C}{{\ensuremath{\mathcal{C}}}}
\newcommand{\complex}{{\ensuremath{\mathbb{C}}}}

\renewcommand{\S}{{\ensuremath{\mathcal{S}}}}
\newcommand{\cS}{{\ensuremath{\overline{\mathcal{S}}}}}
\newcommand{\Hin}{{\ensuremath{\mathcal{H}^{\otimes^n}}}}
\newcommand{\HinA}{{\ensuremath{{L^2(A)}^{\otimes^n}}}}
\newcommand{\E}{{\ensuremath{\mathcal{E}}}}
\newcommand{\D}{{\ensuremath{\mathcal{D}}}}

\newcommand{\biangle}[1]{\langle\langle #1\rangle\rangle}

\title{{\bf Non-Stabilizer Quantum Codes from Abelian Subgroups of the
    Error Group}}

\author{
V.~Arvind, Piyush P Kurur, and K.~R.~Parthasarathy\\
Institute of Mathematical Sciences, C.I.T Campus\\ 
Chennai 600113, India\\
email: {\tt\{arvind,krp,ppk\}@imsc.ernet.in}\\}

\date{}

\begin{document}

\maketitle

\begin{abstract}
  This paper is motivated by the computer-generated nonadditive code
  described in Rains et al \cite{rains97nonadditive}. We describe a theory of
  non-stabilizer codes of which the nonadditive code of Rains et al is
  an example. Furthermore, we give a general strategy of constructing
  good nonstabilizer codes from good stabilizer codes and give some
  explicit constructions and asymptotically good nonstabilizer codes.
  Like in the case of stabilizer codes, we can design fairly efficient
  encoding and decoding procedures.
\end{abstract}

\section{Introduction}

Let $A$ be a finite abelian group with operation denoted by $+$ and
identity 0. We identify $A$ with the alphabet of symbols transmitted
on a classical communication channel. Consider the $n$-fold cartesian
product $A^n$ of copies of $A$. Elements of $A^n$ are called words of
length $n$. A commonly used group is $\{0,1\}$ with addition modulo 2.
Let $\hat{A}$ denote the character group of $A$, the multiplicative
group of all homomorphisms {from} $A$ into the multiplicative group of
complex numbers of modulus unity. For $\a=(a_1,a_2,\ldots,a_n)^T\in
A^n$ we define its weight $w(\bold{a})$ to be $\#\{i\mid a_i\neq 0\}$.
We say that a subgroup $\C_n$ of $A^n$ is a \emph{$t$-error correcting
  group code} if for every non-zero element
$\bold{x}=(x_1,x_2,\ldots,x_n)^T$ in $\C_n$, $w(\bold{x})\geq 2t+1$.
In other words, if messages transmitted through a noisy channel are
encoded into words {from} $\C_n$ and during transmission of a word
errors at the output occur in at most $t$ positions, then the message
can be decoded without any error. There is a vast literature on the
construction of $t$-error correcting group codes and the reader may
find an introduction to this subject and pointers to literature in
\cite{sloan, vLin}.

A broad class of quantum error correcting codes known as stabilizer
codes was introduced by Gottesman \cite{gottes} and Calderbank et al
\cite{cald-shor} (also see \cite{kl2,rains97nonadditive,rains}). To
the best of our knowledge, apart {from} one computer-generated example
of a code proposed by Rains et al \cite{rains97nonadditive}, all known
quantum error-correcting codes are stabilizer codes.

In this paper we develop a theory of \emph{nonstabilizer} codes based
on the Weyl commutation relations. The nonadditive code of Rains et al
\cite{rains97nonadditive} is an instance of our theory and we derive
it directly from the theory. Furthermore, we give a general strategy
of constructing good nonstabilizer codes from good stabilizer codes
and give some explicit constructions and asymptotically good
nonstabilizer codes. For a rich family of nonstabilizer codes, we also
give elegant and efficient encoding circuits. We also give a simple
effective decoding procedure for these nonstabilizer codes.

First we introduce some definitions. We choose and fix an
$M$-dimensional complex Hilbert space $\Hi$ and consider the unit
vectors of $\Hi$ as pure states of a finite level quantum system. If
$A$ is a finite abelian group with $M$ elements and $\{e_x\mid x\in
A\}$ is an orthonormal basis of $\Hi$ indexed by elements of $A$ we
express it in the Dirac notation as $\ket{x}=e_x$.  If
$\bold{x}=(x_1,x_2,\ldots,x_n)^T\in A^n$ is a word of length $n$, we
write
\[
\ket{\x}=\ket{x_1x_2\ldots x_n}=e_{x_1}\otimes e_{x_2}\otimes\ldots\otimes e_{x_n}
\]
where the right-hand side is a product vector in the $n$-fold tensor
product $\Hi^{\otimes^n}$ of $n$ copies of $\Hi$. Thus, with the
chosen orthonormal basis, every word $\x$ in $A^n$ is translated into
a basis state $\ket{\x}$ of $\Hi^{\otimes^n}$.

A \emph{quantum code} is a subspace $\C_n$ of$\Hi^{\otimes^n}$. Note
that a pure state in $\Hi^{\otimes^n}$ described by a unit vector
$\ket{\psi}$ in $\Hi^{\otimes^n}$ has density matrix
$\ket{\psi}\bra{\psi}$. A density matrix $\rho$ in $\Hi^{\otimes^n}$
is a non-negative operator of unit trace. In quantum probability, a
projection operator $E$ in $\Hi^{\otimes^n}$ is interpreted as an
event concerning the quantum system and a density matrix $\rho$ as a
state of the quantum system. The probability of the event $E$ in the
state $\rho$ is given by $\Tr\rho E$. Messages to be transmitted
through a quantum channel are encoded into pure states in
$\Hi^{\otimes^n}$. When a pure state $\ket{\psi}$, or equivalently, a
density matrix $\ket{\psi}\bra{\psi}$ is transmitted the channel
output is hypothesized to be a state of the form
\begin{eqnarray}
\rho=\sum_{i}L_i\ket{\psi}\bra{\psi}L_i^{\dagger}\label{eq-1}
\end{eqnarray}
where the operators $\{L_i\}$ belong to a linear subspace $\A$ of the
algebra of all operators on $\Hi^{\otimes^n}$. The operators $\{L_i\}$
may depend on $\rho$, but in order to ensure that $\rho$ is a density
matrix it is assumed that $\bra{\psi}\sum_i
L^{\dagger}_iL_i\ket{\psi}=1$. By the spectral theorem $\rho$ can
be expressed as
\[
\rho = \sum_{j}p_j\ket{\psi_j}\bra{\psi_j}
\]
where $\psi_j$ is an orthonormal set in $\Hi^{\otimes^n}$ and
$\{p_j\}$ is a probability distribution with $p_j>0$ for each $j$.  In
other words, the output state $\rho$ is not necessarily pure even
though the input state is pure.  The operators $L_i$ are called
\emph{error operators} and the linear space $\A$ {from} which they
come is called the \emph{error space}.

Let $P$ be the projection operator corresponding to a quantum code
$\C_n$. The subspace $\D(P)$ of \emph{error operators detected} by $P$
is defined as
\[
\D(P)=\{L\in\B(\Hin)\mid PLP=c.P\mbox{ for some }c\in\complex\}.
\]

It is evident that there is a complex-valued functional
$\phi:\D(P)\rightarrow\complex$ so that we can write $PLP=\phi(L)P$
for all $L\in\D(P)$.

A finite family $\{M_j\}\subseteq\B(\Hin)$ constitutes a set
of \emph{decoding operators} for the code $\C_n$ and error space $\A$
if the following conditions are satisfied.
\begin{itemize}
\item[(a)] $\sum_j M^{\dagger}_jM_j=I$.
\item[(b)] For any pure state $\ket{\psi} \in C_n$ let $\rho$ be the output
  corresponding to a subset $\{L_i\} \subseteq \A$,
\[
\sum_j M_j\rho M^{\dagger}_j=
\sum_{i,j}M_jL_i\ket{\psi}\bra{\psi}L_i^{\dagger}M_j^{\dagger}
=\ket{\psi}\bra{\psi}.
\]
\end{itemize}

In this case we say that $\C_n$ is an \emph{$\A$-error correcting
quantum code}. 

We have the following fundamental theorem of Knill and Laflamme
\cite{KL} which characterizes the errors that a quantum code can
correct.  It essentially states that errors coming from a family $\A$
of operators can be corrected for a quantum code with projection $P$
if and only if
\[
\{L_1^{\dag}L_2\mid L_1, L_2,\in\A\}\subseteq\D(P).
\]

\begin{theorem}{\rm\cite{KL}}\label{kl-theorem}
  Let $\A$ be a family of operators in $\Hin$ and let $\C_n\subset
  \Hin$ be a quantum code with an orthonormal basis
  $\psi_1,\psi_2,\ldots,\psi_d$. Let $P$ be the projection
  corresponding to the code $\C_n$. Then $\C_n$ is an $\A$-error
  correcting quantum code if and only if
\[
\mathcal{A}^\dag \mathcal{A} = \{L_1^{\dag}L_2\mid L_1, L_2\in\A\}\subseteq\D(P).
\]
\end{theorem}

\begin{remark}
  The proof of the above theorem is constructive and yields the
  decoding operators in terms of $\A$ and the basis
  $\psi_1,\ldots,\psi_d$ of $\C_n$.
\end{remark}

Now we specialize the choice of $\A$. Consider all unitary operators
in $\Hin$ of the form $U=U_1\otimes U_2\otimes\ldots\otimes U_n$ where
each $U_i$ is a unitary operator on $\Hi$ and all but $t$ of the
$U_i$'s are equal to $I$. Such a $U$ when operating on
$\psi=\psi_1\otimes\ldots\otimes \psi_n\in\Hin$ produces $U\ket{\psi}$
which is an $n$-fold tensor product that differs {from} $\psi$ in at
most $t$ places. Denote by $\A_t$ the linear span of all such unitary
operators $U$. A quantum code $\C_n$ is called a \emph{$t$-error
  correcting quantum code} if $\C_n$ is an $\A_t$-correcting quantum
code. 

\section{Quantum Codes and Projections in a Group Algebra}\label{defs}

Let $(A,+)$ be a finite abelian group with $M$ elements and identity
denoted by 0. By the fundamental theorem of finite abelian groups,
$A$ is isomorphic to $\bigoplus_{i=1}^k\Ints_{n_i}$ via the isomorphism
$\tau$. For every $m$, let $\om_m=e^{2\pi i/m}$. Define the
\emph{canonical bicharacter} of the group $A$ as the following
complex-valued function on $A\times A$.
\[
\biangle{a,b}=\prod_{j=1}^k \om_{n_j}^{x_jy_j},\mbox{ where }\\
\tau(a)=(x_1,\ldots,x_k)\mbox{ and }\tau(b)=(y_1,\ldots,y_k).
\]

Notice that for all $a,b,c\in A$ we have
$\biangle{a,b}=\biangle{b,a}$,
$\biangle{a+b,c}=\biangle{a,c}\biangle{b,c}$, and $\biangle{a,b}=1$
for all $b\in A$ if and only if $a=0$. Denote by $\hat{A}$ the
character group of $A$. For each fixed $a\in A$, the bicharacter
$\biangle{a,b}$, as a function of $b$, is a distinct element $\chi_a$
of $\hat{A}$ and the correspondence $a\mapsto \chi_a$ is a group
isomorphism between $A$ and the multiplicative character group
$\hat{A}$.

Denote by $\Hi$ the $M$-dimensional Hilbert space $L^2(A)$ of all
complex-valued functions on $A$, spanned by $\{\ket{x}\}_{x\in A}$
(where the vector $\ket{x}$ denotes the indicator function $1_x$ of
the singleton $\{x\}$).  Define the unitary operators $U_{a}$ and
$V_{a}$ on $\Hi$ for every $a\in A$ by
\[
U_a\ket{x}=\ket{x+a},\hspace{2cm} V_{a}\ket{x}=\biangle{a,x}\ket{x}
\]
where $x\in A$. Then we have

\[
U_aU_b=U_{a+b}~~~~V_aV_b=V_{a+b},\mbox{ and }\\
\biangle{a,b}U_aV_b=V_bU_a\mbox{~~}\forall\mbox{~~} a, b\in A.
\]

These are the Weyl commutation relations between the unitary operators
$U_{a}$ and $V_{a}$ on $\Hi$. The family of operators $\{U_aV_b\mid a,
b\in A\}$ is irreducible. 

The canonical bicharacter on $A$ gives rise to the following
bicharacter on $A^n$.  For two elements $\a=(a_1,\ldots,a_n)$ and
$\b=(b_1,\ldots,b_n)$ in $A^n$, $\biangle{\a,\b}$ is defined as
\[
\biangle{\a,\b}=\prod_{i=1}^n\biangle{a_i,b_i}.
\]

Put $U_{\a}=U_{a_1}\otimes\ldots\otimes U_{a_n}$ and
$V_{\b}=V_{b_1}\otimes\ldots\otimes V_{b_n}$. Then $\{U_{\a}V_{\b}\mid
\a, \b\in A^n\}$ is again an irreducible family of unitary operators
such that $U_{\a}U_{\b}=U_{\a+\b}$ and $V_{\a}V_{\b}=V_{\a+\b}$, and
they satisfy the Weyl commutation relations

\[
\biangle{\a,\b}U_{\a}V_{\b}=V_{\b}U_{\a}\mbox{~~}\forall\mbox{~~}
\a, \b\in A^n.
\]

In the Hilbert space $\B(\Hin)$ of all linear operators on $\Hin$ with
the scalar product $\braket{X}{Y}=\Tr X^{\dagger}Y$ the set
$\{M^{-n/2}U_{\a}V_{\b}\mid \a, \b\in A^n\}$ is an orthonormal basis.
In particular

\[
\Tr~U_{\a}V_{\b}=
\left\{
\begin{array}{cc}
    0&\textrm{if } (\a,\b) \neq (0,0), \\
    M&\textrm{otherwise,}
\end{array}
\right.
\]

The weight $\wt(\a,\b)$ of a pair $(\a,\b)\in A^n\times A^n$ is
defined to be $\#\{i\mid 1\leq i\leq n, (a_i,b_i)\neq (0,0)\}$, where
$\a=(a_1,a_2,\ldots,a_n)$ and $\b=(b_1,\ldots,b_n)$. The
irreducibility of $\{U_{\a}V_{\b}\mid \a, \b\in A^n\}$ implies that
$\{U_{\a}V_{\b}\mid \a, \b\in A^n, \wt(\a,\b)\leq t\}$ spans $\A_t$.
As a result, the Knill-Laflamme theorem for $\A_t$-correcting quantum
codes takes the following form which is easy to derive from
Theorem~\ref{kl-theorem}.

\begin{theorem}\label{kl-2}
  $\C_n\subset L^2(A)^{\otimes^n}$ is a $t$-error correcting quantum
  code if and only if 
\[
\{U_{\a}V_{\b}\mid \a, \b\in A^n, \wt(\a,\b)\leq 2t\}\subseteq\D(P),
\]
where $P$ is the projection corresponding to $\C_n$.
\end{theorem}

Let $N$ be the least positive integer such that $Na = 0$ for all $a
\in A$. Let $\om=e^{2\pi i/N}$, We define the error group $\E$ as
follows.

\begin{definition}
  The \emph{error group} $\E$ is defined as

\[ 
\E = \{\om^iU_{\a}V_{\b}\mid 0\leq i\leq N-1,\a, \b \in A^n\},
\]
with the group operation defined by

\[ 
\om^iU_{\a}V_{\b}\om^jU_{\c}V_{\d}=\om^{i+j}\biangle{\b,\c}U_{\a+\c}V_{\b+\d}.
\]

\end{definition}

Let $\S \leq \E$ be a subgroup of the error group. In the theory of
stabilizer codes we are interested in subspaces of
$L^2(A)^{\otimes^n}$ that are left invariant under the action of $\S$.
It turns out that the invariant subspace is nontrivial if and only if
$\S$ is abelian with the property that $\omega^i I \not \in \S$ for
all $1 \leq i \leq N -1$.

\begin{definition}\label{gottes}
  An abelian subgroup $\S$ of $\E$ is said to be a \emph{Gottesman
    subgroup} of $\E$ if $\omega^i I \not \in \S$ for all $1 \leq i
  \leq N -1$. The \emph{closure} of a Gottesman subgroup $\S$ is the
  abelian subgroup $\cS$ of $\E$ defined as
\[
\cS=\{\om^i g\mid g\in\S,\ 0\leq i \leq N -1\}.
\]
\end{definition}

\begin{remark}
  For any Gottesman subgroup $\S$ of the error group $\E$ the element
  $\omega^i U_{\a} V_{\b}\in\S$ for at most one $i: 0\leq i\leq N-1$.
\end{remark}

Let $\mathbb{C}[\S]$ be the group algebra of formal sums $\sum_{s \in
  \S} T_s s$, $T_s \in \mathbb{C}$ where the plus and product
(convolution) are defined as follows

\[
T+T'=\sum_{s \in \E} (T_s+T'_s) s,
\]

\[
T'*T''=T,\mbox{\rm where }T_s=\sum_{g\in\E} T'_g T''_{g^{-1}s}.
\]

Since any Gottesman subgroup $\S$ is a set of linearly independent
elements of $\B(\Hin)$, the identity map is a natural injective linear
embedding from $\mathbb{C}[\S]$ into $\B(\Hin)$. In other words,
$\mathbb{C}[\S]$ is a subalgebra of $\B(\Hin)$ under operator addition
and composition, where the convolution operation ``$*$'' in $\mathbb{C}[\S]$
coincides with composition operator.

Every subspace (i.e.\ quantum code) in $\Hin$ is defined by its
corresponding projection operator in $\B(\Hin)$. In this paper we are
interested in the projection operators in $\mathbb{C}[\S]$ for a Gottesman
subgroup $\S$ of $\E$. In the following easy proposition we
characterize the elements in $\mathbb{C}[\S]$ which are projection operators
in $\Hin$. Then our goal will be to seek for projections whose range
is a good $t$-error correcting quantum code.

\begin{proposition}
  The element $T\in\mathbb{C}[\S]$ is a projection operator on $\Hin$ if and
  only if $\overline{T_s}=T_{s^{-1}}$ and $T*T=T$.
\end{proposition}

We use the Fourier transform over $\mathbb{C}[\S]$ and the above proposition to
describe projection operators in $\mathbb{C}[\S]$. We recall the Fourier
transform and some of its properties.

Let $\S$ be an abelian group and let $\hat{\S}$ denote the character
group of $\S$. For each $s\in \S$ we can associate the element
$\sum\alpha_g g$ in the algebra $\mathbb{C}[\S]$ where, $\alpha_s = 1$ and
$\alpha_g=0$ for $g\neq s\in\S$. Similarly, to $\chi\in\hat{S}$ we
associate the element $\sum\chi(g)g$ of $\mathbb{C}[\S]$. Fix an isomorphism
$s \mapsto \chi_s$ between the groups $\S$ and $\hat{\S}$. The Fourier
transform over $\mathbb{C}[\S]$ is now defined as follows.

\begin{definition}[Fourier Transform]
  The linear transformation that maps $s \in \mathbb{C}[\S]$ to $\chi_s \in
  \mathbb{C}[\S]$ is called the \emph{Fourier transform} over $\mathbb{C}[\S]$. The
  Fourier transform $\hat{T}$ of $T$ in $\mathbb{C}[\S]$ is given by the
  formula
  \[ \hat{T}_u = \sum_{s \in \S} \chi_u(s) T_s.\]
\end{definition}

The \emph{inverse Fourier transform} is given by the following
formula:

\[
T_s= \frac{1}{\#\S} \sum_{u \in \S} \overline{\chi}_u(s) \hat{T}_u .
\]

Let $TT'$ denote the component-wise product of $T, T'\in\mathbb{C}[\S]$. I.e.

\[
TT' =\sum_{s\in\S}T_sT'_s s.
\]

We now recall some useful properties of the Fourier transform.

\begin{eqnarray}
\nonumber
  \widehat{{T_1*T_2}} &=& \hat{T_1}\hat{T_2},~~~~~~ T_1, T_2\in\mathbb{C}[\S]. \\
\nonumber
  \widehat{g T} &=& \chi_g \hat{T},~~~~~~ g\in\S,~~~ T\in\mathbb{C}[\S].
\end{eqnarray}

We can characterize projection operators in $\mathbb{C}[\S]$ using the
Fourier transform.

\begin{theorem}\label{nstab-code}
  An element $T \in \mathbb{C}[\S]$ is a projection (and hence a code) iff
  $\hat{T} = 1_B = \sum_{g \in B} g$ for some subset $B$ of $\S$.
\end{theorem}

\begin{proof}
  $T$ is a projection iff $T*T = T$ and $T^\dagger = T$. On taking
  Fourier transforms on both sides of the equation $T*T = T$ we get
\[
\hat{T} \hat{T} = \hat{T}.
\]

This implies $\hat{T}_s^2=\hat{T}_s$ for every $s\in\S$. Thus,
$\hat{T}_s\in\{0,1\}$ for every $s\in\S$ which gives the desired result.

Conversely, note that if $\hat{T} = 1_B$ then by inverse Fourier
transform we get

\[
T = \frac{1}{\#\S}\sum_{s \in \S} \sum_{u \in B} \overline{\chi_u}(s) s.
\] 

From the above equation it is clear that the condition $T^\dagger = T$
is automatically satisfied.
\end{proof}

From Theorem~\ref{nstab-code} it is clear that the code defined by a
projection $T$ in $\mathbb{C}[\S]$ is completely specified by the subset $B$
of $\S$. We will call $B$ the {\em Fourier description} of the quantum
code and denote the corresponding projection by $P(B)$. 

Next, we give a formula for the dimension of a quantum code defined by
a projection $T$ in $\mathbb{C}[\S]$.

\begin{lemma}{\label{dimlemma}}
  Let $\C_n\subseteq\Hin$ be a quantum code defined by a projection
  $T$ in $\mathbb{C}[\S]$ with Fourier description $B \subseteq \S$. Then
  $\C_n\neq 0$ if and only if $\om^iI\not\in\S$ for $i: 1\leq i\leq
  N-1$ (i.e. $\S$ is a Gottesman subgroup), and if $\C_n\neq 0$ the
  dimension of the code $\C_n$ is given by
\[ 
\dim(\C_n) =\frac{\#A^n \#B}{\#\S}.
\]
\end{lemma}

\begin{proof}
  If $B$ is the Fourier description of the code then the projection
  corresponding to the code is given by \[ P =\frac{1}{\#\S} \sum_{s
    \in \S} \sum_{u \in B} \overline{\chi_u}(s) s. \] The dimension is
  given by $\Tr(P)$.  Observe that $\Tr(U_{\a}V_{\b}) = 0$ if
  $U_{\a}V_{\b}\neq I$, and $\Tr(\om^iI)=\om^i\#A^n$, for
  $\om^iI\in\S$. Now, since $\om$ is a nontrivial root of unity,
  $\sum_{\om^iI\in\S} \om^i \# A =0 $ if $\om^iI\in\S$ for some $i$
  such that $1\leq i\leq N-1$. Thus, we have $\Tr(P)=0$ if
  $\om^iI\in\S$ for some $i:~1\leq i\leq N-1$, and otherwise
  $\Tr(P)=\frac{\#B}{\#\S}\#A^n$. This proves the lemma.
\end{proof}

\begin{proposition}
  Let $g = \omega^i U_{\a} V_\b$ and $h = \omega^j U_\c V_\d$ be
  elements in the error group $\E$. Then

  \[
  ghg^{-1}h^{-1} = \gamma(g,h) I
  \]
  where $\gamma(g,h) = \biangle{b,c} \overline{\biangle{a,d}}$. In particular
  we have 

  \[
  \gamma(g_1g_2,h) = \gamma(g_1,h)\gamma(g_2,h)
  \]

  \[
  \gamma(g,h_1 h_2) = \gamma(g,h_1) \gamma(g,h_2)
  \]
  and
  \[
  \gamma(g,h) = \overline{\gamma(h,g)}
  \]
\end{proposition}
\begin{proof} 
  Straightforward from the Weyl commutation relations.
\end{proof}
\begin{remark}\label{gammagremark}
  When $h \in \E$ is fixed and $s$ varies in $\S$ the map $s \mapsto
  \gamma(s,h)$ is a character of $\S$ which we will denote by
  $\gamma_h$.
\end{remark}

Let $\C_n\subseteq\Hin$ be a code with Fourier description $B$. In the
next theorem we derive a condition on $B$ such that $\C_n$ is a
$t$-error correcting quantum code. We introduce a convenient notation:
For $g=\om^iU_{\a}V_{\b}$ in the error group $\E$, let $\wt(g)$ denote
the number $\wt(\a,\b)$.

\begin{theorem}{\label{distdtheorem}}
  A quantum code $\C_n\subseteq\Hin$ with Fourier description $B$ is
  $t$-error correcting iff the following two conditions hold.
  
\begin{enumerate}
  
\item 
For each $g \in \S$ such that $\wt(g)\leq 2t$
\[
\chi_g( u_1^{-1} u_2) = 1 \textrm{ for all } u_1,u_2 \in B. 
\]
(i.e. every $u \in B$ is in the same coset of the kernel of $\chi_g$.)

\item 
For each $g \in \E \setminus \S$ such that $\wt(g)\leq 2t$, and for 
every $u \in B^{-1} B$ we have
\[
\sum_{s \in \S} \gamma_g(s)\chi_u(s) = 0.
\]
(i.e.\ the character $\gamma_g$ is different from $\chi_{u^{-1}}$ for every
$u\in B^{-1} B$.)

\end{enumerate}
\end{theorem}

\begin{proof}
  
  Let $\C_n$ be a $t$-error correcting code and let $T = \sum_{s\in\S}
  T_s s$ be the corresponding projection. Recall that if $B$ is the
  Fourier description for the code then
\[
T_s = \frac{1}{\# \S} \sum_{u \in B} \overline{\chi}_u(s).
\]
  
By Theorem~\ref{kl-theorem}, $\C_n$ is $t$-error correcting if and
only if there is a scalar-valued function $\phi$ such that for every
$g\in\E$ with $\wt(g)\leq 2t$
\[
TgT = \phi(g) T.
\]

This is equivalent to the following condition.

\begin{eqnarray*}\label{eq-3}
\left( \sum_{s_1 \in \S} T_{s_1} s_1 \right) g 
\left( \sum_{s_2 \in \S} T_{s_2} s_2 \right) = \phi(g)\sum_{s\in\S}T_s s.
\end{eqnarray*}

\paragraph{{\em Case 1}~~~ $g \in \S$ :}

In this case Equation~\ref{eq-3} yields $gT*T =\phi(g)T$. Taking Fourier
transform on both sides we get 

\[ \chi_g \hat{T} \hat{T} = \phi(g)\hat{T}.\] 

Since $\hat{T} = 1_B$, we have $\chi_g(u) = \phi(g)$ for all $u \in
B$.  Thus, $\chi_g$ is constant on $B$ for every $g \in \S$ such that
$\wt(g)\leq 2t$. This is true precisely when $B$ is contained in some
coset of the kernel of $\chi_g$.

\paragraph{{\em Case 2} $g\not\in\S$ :}

In this case the Knill-Laflamme condition takes the following form

\[
\sum_{s\in\S}
\sum_{s_1s_2 = s} T_{s_1}T_{s_2} \gamma(s_1,g) g s_1 s_2 = \phi(g)
\sum_{s \in \S} T_s s.
\]

Since the operators on the two sides of the above equation have
disjoint support, each side of the equation should to be $0$. Consequently,
$\phi(g) = 0$ and for all $s \in \S$
\[
\sum_{s_1s_2 =s} T_{s_1} T_{s_2} \gamma(s_1,g) = 0.
\] 

This yields

\[
\sum_{s_1s_2 =s} \left(\sum_{u_1 \in B}\overline{\chi}_{u_1}(s_1)
\right) \left(\sum_{u_2 \in B}{\overline{\chi}_{u_2}}(s_2)\right)
\gamma(s_1,g) = 0.
\]

On simplification we get
\[
\sum_{u_1,u_2 \in B} \sum_{s_1 s_2 = s } \gamma_g(s_1)
\overline{\chi}_{u_1}(s_1) \overline{\chi}_{u_2}(s_2) = 0,
\]
which gives 

\begin{equation}\label{eqn1}
\sum_{u_1,u_2 \in B} \overline{\chi}_{u_2}(s) \sum_{s_1 \in \S}
(\gamma_g\chi_{u_2} \overline{\chi}_{u_1})(s_1) = 0.
\end{equation}
 
Note that the inner summation in equation~\ref{eqn1} is summing up of
a character of $\S$, namely, $\gamma_g \chi_{u_1^{-1}} \chi_{u_2}$,
over the whole group $\S$. Therefore, the inner summation evaluates to
either $0$ or $\# \S$. Hence, the necessary and sufficient condition
for equation~\ref{eqn1} to hold is
\[
\sum_{s \in \S}\gamma_g(s)\chi_u(s) = 0~~~~~\forall u \in B^{-1}B.
\]

\end{proof}

\begin{remark}
  If the Fourier description of a code, $B$, is a subgroup of $\S$
  then the code is actually a stabilizer code with stabilizer group
  $B^\perp$, where $B^\perp$ is the annihilator of $B$ in $\S$ defined
  by
  \[
  B^\perp = \{ a \in \S : \forall b \in B\ \biangle{a,b} = 0 \}.
  \]
  In particular if we set $B=\{I\}$, where $I$ is the identity
  element, the code $\C_n$ with Fourier description $B$ is the
  stabilizer quantum code:
  \[
  \C_n = \{\ket{\psi}\mid s\ket{\psi}=\ket{\psi} \forall s\in\S\}.
  \]
  Thus the stabilizer codes of \cite{gottes} are a subclass of the class
  of codes defined in this paper.
\end{remark}

At this point we recall some useful facts from the theory of
stabilizer codes as developed in \cite{AP02},

Let $\S$ be a Gottesman subgroup of $\E$
\[
\C(\S)=\{\psi\in\HinA\mid U\psi=\psi\mbox{~~}\forall\mbox{~~} U\in \S\}.
\]

Let $Z(\S)$ denote the centralizer of $\S$ in $\E$, that is,

\[
Z(\S) = \{U\in\E\mid UU'=U'U \mbox{~~}\forall\mbox{~~} U'\in \S\}.
\]

For Gottesman subgroup $\S$ recall that the closure of $\S$ (denoted by
$\overline{\S}$) is defined as
\[
\overline{\S} = \{ \omega^i U_x V_y : 0 \leq i \leq N-1 \textrm{ and }
 \exists j\ 0 \leq j \leq N-1 \textrm{ and } \omega^j U_x V_y \in \S \}.
\]

\begin{theorem}\label{kl-3}{\rm\cite{AP02}}
  Let $\S$ be an Gottesman subgroup of the error group $\E$ and $\cS$
  be the closure of $\S$. Then $\C(\S)$ is a $t$-error correcting
  quantum code if $\wt(\a,\b)>2t$ for each $\om^iU_{\a}V_{\b}\in
  Z(\S)\setminus\cS$.
\end{theorem}

We introduce a useful notation for describing quantum stabilizer
codes. Let $\S$ be a Gottesman subgroup of $\E$ with centralizer
$Z(\S)$. The \emph{minimum distance} $d(\S)$ is defined to be the
minimum of
\[
\{\wt(\a,\b)\mid \om^iU_{\a}V_{\b}\in Z(\S)\setminus\cS\}.
\]

When $A$ is the additive abelian group of the finite field $\F_q$ we
define an $\brac{n,k,d}{q}$ quantum stabilizer code to be a
$q^k$-dimensional subspace $\C(\S)$ of $L^2(\F_q)^{\otimes^n}$, where
$\S$ is a Gottesman subgroup of $\E$ with $d(\S)\geq d$ and cardinality
$q^{n-k}$.

By Theorem~\ref{kl-3} it follows that an $\brac{n,k,d}{q}$ quantum
stabilizer code is a $\floor{(d-1)/2}$-error correcting quantum code.

\section{Nonstabilizer codes over finite fields}\label{finitefieldsection}

We focus our attention to the case when the abelian group $A$ is the
field $ \GF{q}$. Recall that the additive group $\GF[n]{q}$ is a
vector space over $\mathbb{F}_q$. If $\omega$ is a nontrivial
character then the characters of the additive group $\GF[n]{q}$ is the
set $\{ \omega_{\bf a} | {\bf a} \in \GF[n]{q} \}$ where $\omega_{\bf
  a}({\bf b}) = \omega( {\bf a} . {\bf b})$, $\a . \b = \sum a_i
b_i$.

The set of operators $\{ \omega_p^i U_{\bf a} V_{\bf b} | {\bf a},{\bf
  b} \in \GF[n]{q} \}$, $\omega_p$ is a $p^{th}$ root of unity, where
$p$ is the characteristic of the field $\GF{q}$ forms an irreducible
representation of associated error group.  It is shown in \cite{AP02}
that any Gottesman subgroup $\S$ is of the following form

\begin{eqnarray*}\label{eq-lm}
\{ \omega(\rho({\a}))U_{L{\a}}V_{M{\a}}\mid {\bf a}\in\GF[r]{q}\},
\end{eqnarray*}
where $L$ and $M$ are $n \times r$ matrices over $\GF{q}$ such that $L^{T}M$ is
symmetric and $\rho(.)$ satisfies the condition that
\[
\rho({\v_1} + {\v_2}) - \rho({\v_1}) - \rho({\v_2}) = {\v_2}^{T}L^{T} M{\v_1}.
\]

Our goal in this section is to seek for nonstabilizer codes with the
help of Theorem~\ref{distdtheorem}.

\begin{definition}
A Gottesman subgroup $\S$ of the error group $\E$ is said to be
\emph{$d$-pure} if $\wt(g)\geq d$ for every $g\in Z(\S)$.
\end{definition}

By the theory of stabilizer codes, it follows that the corresponding
stabilizer code $\C(\S)=\{\psi\in\HinA\mid
U\psi=\psi\mbox{~~}\forall\mbox{~~} U\in \S\}$ is a
$\floor{(d-1)/2}$-error correcting quantum code.

More precisely, our aim is to start with the stabilizer code $\C(\S)$
of distance $d$, and use Theorem~\ref{distdtheorem} to construct
nonstabilizer codes of the same distance but larger dimension.

Observe that if $\S$ is a $2t$-pure Gottesman subgroup of $\E$, the
first condition in Theorem~\ref{distdtheorem} is vacuously true. Thus,
we only need to ensure that the second condition in
Theorem~\ref{distdtheorem} is satisfied.  For a $d$-pure Gottesman
code we define the \emph{forbidden set} as follows

\begin{definition}
  Let $\S$ be a $d$-pure Gottesman subgroup of the error group $\E$.
  We define the \emph{$d$-forbidden subset} of $\S$, denoted by
  $\mathcal{F}_d(\S)$, to be the subset
  \[ 
  \mathcal{F}_d(\S) = \{ u \in \S : \exists g \in \E \setminus \S\  wt(g) < d 
  \textrm{  and }\sum_{s \in S}(\gamma_g \chi_u)(s) = \# \S \}.
  \]
\end{definition}

We have the following theorem that is straight forward consequence of
Theorem~\ref{distdtheorem}.

\begin{theorem}\label{puredistdtheorem}
Let $\S$ be a $d$-pure Gottesman subgroup of the error group $\E$ then 
$B \subseteq \S$ is the Fourier description of a distance $d$ code \emph{iff}
$B^{-1}B \cap \mathcal{F}_d(\S)$ is empty.
\end{theorem}

Let $s_{\a}$ denote $\omega_p(\rho(a)) U_{L\a} V_{M\a}\in\S$. Observe
that $\tau: s_{\a}\mapsto \a$ is a group isomorphism from $\S$ to
$\GF[r]{q}$, and $\chi_{s_{\a}}\mapsto \om_{\a}$ is an isomorphism
from $\widehat{\S}$ to $\widehat{\GF[r]{q}}$.
    
Let $g =\om^i U_\x V_\y\not\in\S$ with $\wt(g)\leq 2t$. By applying
the Weyl commutation relations we get $\gamma_g(s_{\a})=
\omega(\a^{T}M^T \x - \a^{T}L^{T}\y) = \omega_\a( M^T \x - L^T \y)$.
We have the following theorem.

\begin{theorem}\label{forbiddensettheorem}
  If $\S = \{ \omega(\rho(\a)) U_{L\a} V_{M\a} : a \in \GF[r]{q} \}$
  is $d$-pure Gottesman subgroup of the error group $\E$ over
  $\GF[n]{q}$. The $d$-forbidden subset of $\S$ is given by
  \[
  \mathcal{F}_d(\S) = \{ s_{\bf u}: \exists \x,\y \in \GF[n]{q}\ wt(\x,\y) < d
  \textrm{ and } {\bf u} = L^T \y - M^T \x \}
  \]
\end{theorem}
\begin{proof}
  \[
  \sum_{s \in \S} \gamma_g(s) \chi_u(s) = \sum_{\a \in \GF[r]{q}}
  \omega_a({\bf u } + M^T \x - L^T \y)
  \]
  Note that the right hand side of the equation is the sum over all
  character of $\GF[r]{q}$ and hence is nonzero iff $ {\bf u} + M^T \x
  - L^T \y= 0$. Hence
  \[
  \mathcal{F}_d(\S) = \{ s_{\bf u}: \exists \x,\y \in \GF[n]{q}\ wt(\x,\y) < d
  \textrm{ and } {\bf u} =L^T\y -  M^T \x  \}
  \]
\end{proof}

\begin{remark}\label{forbiddensetremark}
  In the above setting we will call the set 
  
  \[
  F_d(\S) = \{ u: \exists \x,\y \in \GF[n]{q}\ wt(\x,\y) < d
  \textrm{ and } {\bf u} = L^T \y - M^T \x \}
  \] 
  the forbidden set. Note that $\mathcal{F}_d(\S) = \{ s_{\bf u} : {
    \bf u } \in F_d(\S)\}$.
\end{remark}
\section{Bounds on the dimension of codes}

We now give upper and lower bounds on the dimension of nonstabilizer
codes built from pure Gottesman subgroups of $\E$. Let the encoding
space be $\HinA$, and $N(n,q,d)$ denote the number $\sum_{i=0}^{d}
{\left( \begin{array}{c} n \\ i \end{array} \right)} (q^2 -1)^i$. We have the following upper bound on the
dimension of the code.

\begin{theorem}
  Let $\C_n\subseteq\HinA$ be a $d$-error correcting quantum code such
  that its corresponding projection $P$ has support in a $2d+1$-pure
  Gottesman subgroup $\S$ of the error group $\E$. Let $B$ be the 
  Fourier description of $\C_n$. Then the dimension of the code $\C_n$
  satisfies the inequality 
\[ 
\dim(\C_n) \leq \frac{\#A^n}{N\left(n,\#A,d\right)}. 
\]
\end{theorem}

\begin{proof}
  Since $\S$ is $2d+1$-pure, $\{g\in\E\mid \wt(g)\leq 2d\}\subseteq
  \E\setminus\S$. By Theorem~\ref{distdtheorem} we have:

\[ 
PgP = 0
\] 
for all $g\in\E$ such that $\wt(g)\leq 2d$. Let $P_g$ denote the
projection $g^{-1}Pg$. The range of $P_g$ has dimension $\dim(\C_n)$
for every $g\in\E$. Furthermore, for all $g_1, g_2\in\E$ such that
$\wt(g_1)\leq d$ and $\wt(g_2)\leq d$, we have 

\[ 
P_{g_1} P_{g_2} = g_1^{-1} P g_1 g_2^{-1} P g_2 = 0,
\] 
since $\wt(g_1 g_2)\leq 2d$ implies $P g_1 g_2 P = 0$. Thus,
$\{P_g\mid g\in\E, \wt(g)\leq d\}$ is a collection of mutually
orthogonal projections in $\HinA$. Furthermore, the range of each
$P_g$ is $\dim(\C_n)$. Since there are $N(n,\#A,d)$ elements $g$ in
$\E$ with $\wt(g)\leq d$, it follows by adding dimensions that

\[
N\left(n,\#A,d\right)~\dim(\C_n) \leq \dim{\HinA} = \#A^n.
\] 
\end{proof}

We now show a lower bound for the code dimension for codes satisfying
the conditions of Theorem~\ref{forbiddensettheorem}.

\begin{theorem}\label{lowerbound}
  Let $A=\F^n_q$, and $\S=\{ \omega(\rho(\a)) U_{L \a} V_{M \a} | \a
  \in \GF[r]{q}\}$ be a $2d+1$-pure Gottesman subgroup of the error
  group $\E$ for the encoding space $\HinA$. Then there is a $d$-error
  correcting code $\C_n$ such that its corresponding projection has
  support in $\S$ and
\[
\dim(C_n)\geq \frac{q^n}{N(n,q,2d)}.
\]
\end{theorem}

\begin{proof}
  For the $2d+1$-pure Gottesman subgroup $\S=\{ \omega(\rho(\a)) U_{L
    \a} V_{M \a} | \a \in \GF[r]{q}\}$, let $X$ be the corresponding
  forbidden set. By Theorem~\ref{forbiddensettheorem}, $X$ is the
  image of the set $\{ (\x,\y) | \x,\y\in \GF[n]{q}, wt(\x,\y) \leq
  d-1\}$ under the map $(\x,\y) \mapsto L^T \y - M^T \x$. The number of
  $\x,\y \in \GF[n]{q}$ such that $\wt(\x,\y)\leq 2d$ is $N(n,q,2d)$.
  Hence $\#X\leq N(n,q,2d)$. We prove the existence of the code $\C_n$
  by constructing its Fourier description $B\subseteq \S$ using the
  following ``greedy'' strategy to pick elements from $\S$:

  \begin{enumerate}
  \item initially, let $B = \{ \}$ and let $A = \S$.
  \item Pick any $\u \in A$ and include in $B$.
  \item Remove from $A$ all elements $\v$ such that difference $\u
   -\v$ is in $X$, where $u$ is the element picked in the previous
    step.
  \item If $A=\emptyset$ stop. Otherwise, return to Step 2.
  \end{enumerate}
  Note that this strategy will eliminate at most $\#X$ elements from
  $A$ every time we include a new element in $B$. Thus, the number of
  elements picked into $B$ will be at least
  $\floor{\frac{\#\S}{\#X}}\geq \frac{\#\S}{N(n,q,2d)}$. Applying
  Lemma~\ref{dimlemma} yields the desired lower bound.
\end{proof}

We can now easily argue about the existence of asymptotically good
nonstabilizer codes. The following theorem is a paraphrase of a result
we proved in \cite{AP02} about the existence of $d$-pure maximal
Gottesman subgroups of the error group $\E$ for large $d$, for encoding
space $\HinA$ for large $n$ ($A=\F_2$).

We first need the following technical definition.

\begin{definition}\label{good}{\rm\cite{AP02}}
  An $n\times n$ matrix $R$ over $\F_2$ is said to be
  $\alpha$-\emph{good} if the following conditions are true.
\begin{enumerate}
\item[(i)] The sum of every $\floor{\alpha n}$ columns of $R$ has weight at
  least $\alpha n$.
\item[(ii)] The sum of every $\floor{\alpha n}$ rows of $R$ has weight at
  least $\alpha n$.
\item[(iii)] The sum of every $\floor{\alpha n}$ columns of $R$ has weight at
  most $(1-\alpha)n$.
\item[(iv)] The sum of every $\floor{\alpha n}$ rows of $R$ has weight at
  most $(1-\alpha)n$.
\end{enumerate}
\end{definition}

It is shown in \cite{AP02} that there is a constant $\alpha>0$ and a
corresponding positive integer $n_{\alpha}$ such that

\[
\Pr[R\mbox{ is $\alpha$-good }]>0.
\]
 
\begin{theorem}\label{exists}{\rm\cite{AP02}}
  For $0<\alpha<1$, suppose $R$ is an $n\times n$ $\alpha$-good matrix
  over $\F_2$.  Let $L$ be the following $2n\times 2n$ symmetric
  matrix over $\F_2$:
\[
\left( \begin{array}{cc}
       0 & R \\
       R^T & 0 
       \end{array}   
\right)
\]
If we write $L=D+D^T$, where $D$ is the upper triangular matrix with
zeros on the principal diagonal, then $\S=\{\w(\a^TD\a)U_{\a}V_{L\a
  -\b}\mid \a\in C, \b\in C^{\perp}\}$, is an $\floor{\alpha n}$-pure
maximal Gottesman subgroup $\S$ of the error group $\E$.
\end{theorem}

Now, applying Theorem~\ref{lowerbound} we immediately get the following
family of asymptotically good nonstabilizer codes.

\begin{corollary}\label{asymp-good}
  For $0<\alpha<1$, suppose $R$ is an $n\times n$ $\alpha$-good matrix
  over $\F_2$ and $\S$ is the $\floor{\alpha n}$-pure maximal Gottesman
  subgroup $\S$ of the error group $\E$ (defined in the above
  theorem).  Then there is an $\floor{\alpha n-1/2}$-error correcting
  quantum code of dimension $\frac{2^n}{N(n,2,\floor{\alpha n})}$,
  whose projection has support in $\S$.
\end{corollary}

\section{Explicit construction of Non-stabilizer codes}\label{explicit}

We now give an explicit construction of a family of distance $2$ code.
Recall that any abelian group of the error group is of the form
\[
\S = \{ \omega(\rho(\a)) U_{L\a}V_{M\a} : \a \in \GF[r]{q} \}
\]
where $L$ and $M$ are $r \times n$ matrices over \GF{q} such that $L^T
M$ is symmetric and $\rho$ satisfies the condition
\[
\rho(\a_1 + \a_2 ) - \rho(\a_1) - \rho(\a_2) = \a_1^T L^T M \a_2.
\] 

Given an odd integer $n = 2m + 1$, we give the explicit construction
of a $((n,1+n(q-1),2))_q$ code. Note that if $q =2$ and $n=5$ we get a
$((5,6,2))_2$ code. In \cite{rains97nonadditive} a $((5,6,2))_2$ code
is given which is generated by a computer search. They have also shown
that for distance 2 this is the best possible code. We also show that
there is a code of dimension greater than $\left\lceil
  \frac{q^n}{n(q^2-1)} \right\rceil$.

Let ${\bf x} \in \GF[n]{q}$ is all zeros except at positions $m+1$ and
$m+2$ where it is $1$. Define the matrices $S$ and $L$ as follows

\[
S = \left( \begin{array}{c}
                    {\bf x}\\
                    \sigma{\bf x}\\
                    \vdots\\
                    \sigma^i{\bf x}\\
                    \vdots\\
                    \sigma^{n-1}{\bf x}
                  \end{array}
                \right)\  L =
                \left( 
                 \begin{array}{c|c}
                    &  0\\
                    I_{n-1}   & \vdots \\
                    & 0 \\
                    \hline\\
                    -1 \ldots -1     & 0 
                  \end{array}
                \right)
\] 
where $\sigma$ is the cyclic shift on $n$ elements and $I_{n-1}$ is
the $n-1 \times n-1$ identity matrix. Let $J$ be the $n\times n $
matrix, all of whose entries are $1 \in \GF{q}$. Note that $JL = 0$
and hence $M^TL = L^T S L$ is symmetric. As a result $L$ and $M$ gives
rise to a Gottesman subgroup
\[
\S = \{ \omega(\rho(a)) U_{L\a} V_{M\a} : \a \in \GF[n]{q} \}
\]

Let $\e_i, 0 \leq i \leq n - 1$ be the standard basis for $\GF[n]{q}$.
$\e_i$ is the vector with a $1$ in the $i^{th}$ position and $\e_0$ is
the vector with a $1$ at the $n^{th}$ position. Let $\underline{1} =
(1,1,\ldots,1)$ then we have the following observation.
\begin{observation}
  \[
  S \e_j  = \e_{j + m} + \e_{j + m + 1}\ ( \textrm{ index addition mod n} ).
  \]

  \[
  L^T \e_j = \left\{
    \begin{array}{cc}
      \e_0 - \underline{1} & j = 0\\
      \e_j & otherwise\\
    \end{array}
  \right.
\]

\[
M^T \e_j = \left\{
  \begin{array}{cc}
    \e_{j+m} + \e_{j+m+1} + \underline{1} & \textrm{if } j+m \not \equiv 0 \textrm{ and } 
    j+m+1 \not \equiv  0\ (\textrm{mod n})\\
    \e_{0} + \e_1  & j+m \equiv 0\ (\textrm{mod n})\\
    \e_0 + \e_{n-1} & j+m+1 \equiv 0\ (\textrm{mod n})
    \end{array}
    \right.
\]
\end{observation}
Note that for $L$ and $M$ defined as above $\S$ will be a maximal abelian subgroup (because its 
cardinality is $2^n$) and is $2-pure$. More over the 2-forbidden set is given by
\[
F_2 = \{ L^T \y - M^T \x : wt(x,y) = 1 \} = \{ a M^T \e_j + b L^T \e_j : (a,b) \ne (0,0);\ a,b \in \GF{q}
                                                                           1 \leq j \leq n \}
\]
    
We have the following asymptotic result.

\begin{theorem}
  Let $n = 2m +1$ be an odd integer. There exists a
  $((n,\left\lceil\frac{q^n}{n(q^2 -1)}\right\rceil,2))_q$ quantum code.
\end{theorem}
\begin{proof}
  $\#F_2 = n (q^2 - 1)$. Applying a greedy algorithm similar to the
  one in theorem~\ref{lowerbound} we get the required result.
\end{proof}

Consider the subset $B$ of $\S$ defined as
\[
B = \left\{ \underline{0} \right\} \cup \left\{ \alpha \e_0 : \alpha \in \GF[*]{q} \right\} \cup
\left\{ \e_0 + \alpha\left( \sum_{i=1}^{n-1} \e_i\right) - \e_j : \alpha \in
  \GF[*]{q};\ j = 1,2,\dots n-1 \right\}.
\]
We have $\# B = 1 + n(q-1)$. We also have the following theorem
\begin{theorem}
  The set $B$ as defined above is the Fourier description of a $((n,1 + n(q-1), 2))_q$ code.
\end{theorem}
\begin{proof}
  Let $D = (B - B ) \setminus \{ \underline{0} \}$.  Since $F_2$ does
  not contain the zero vector it is sufficient to prove that $F_2 \cap
  D$ is empty. Let $\u_i = \left(\sum_{j=1}^{n-1} \e_j\right) - \e_i, 1 \leq i \leq
  n-1$.  Now $D = A_1 \cup A_2 \cup A_3 \cup A_4$ where
  \[
  \begin{array}{ccl}
    A_1 &=& \{ \alpha \e_0 , \alpha \ne 0 \} \\
    A_2 &=& \{ \e_0 + \alpha \u_i : \alpha \ne 0 \} \\
    A_3 &=& \{ \alpha \e_0 + \beta \u_i : \beta \ne 0;\ \alpha \ne -1 ;\ 1\leq i \leq n - 1 \}\\
    A_4 &=& \{ \alpha \u_i + \beta \u_j: \alpha,\beta \ne 0;\ 1 \leq i,j \leq n - 1;\
     \alpha \ne \beta \textrm{ or } i \ne j\}
  \end{array}
  \]
  where $\alpha,\beta \in \GF{q}$. The elements of the forbidden set
  $F_2$ are given by
  \[
  \begin{array}{ccl}
    R_1 &=& \{ a (\e_m + \e_{m+1}) + b \e_0 + (a-b)\underline{1} : (a,b) \ne (0,0) \}\\
    R_2 &=& \{ a ( \e_0 + \e_1 ) + b \e_{m+1} : (a,b) \ne (0,0) \}\\
    R_3 &=& \{ a (\e_0 + \e_{n-1} ) + b \e_m : (a,b) \neq (0,0) \}\\
    R_4 &=& \{ a (\e_{j+m} + \e_{j+m+1}) + b \e_j + a \underline{1} : (a,b) \ne (0,0);\
                                                                    j \ne 0, j\ne m, j \ne m+1 \}
  \end{array}
  \]
  Now it can be verified that $A_i \cap R_j$ is empty for every $1 \leq i,j \leq 4$. 
\end{proof}

\section{Examples of 1-error correcting nonstabilizer codes}

In this section we give explicit constructions of a $((33,155,3))$
code and a $((15,8,3))$ code. The codes we construct will be over the
field $\GF{2}$.

Let $n=2m+1$ be any odd integer.  Let ${\bf x }$ be the vector in
$\GF[n]{2}$ with zeros at all positions except $m+1$ and $m+2$. As in the previous section, let
\[
S = \left( \begin{array}{c}
                    {\bf x}\\
                    \sigma{\bf x}\\
                    \vdots\\
                    \sigma^i{\bf x}\\
                    \vdots\\
                    \sigma^{n-1}{\bf x}
                  \end{array}
                \right)\  L =
                \left( 
                 \begin{array}{c|c}
                    &  0\\
                    I_{n-1}   & \vdots \\
                    & 0 \\
                    \hline\\
                    1 \ldots 1     & 0 
                  \end{array}
                \right),
\] 

where $\sigma$ is the cyclic shift. Let $J$ be the $n \times n$ 
matrix all of whose entries are 1's.  Recall that the generalized
Laflamme code is the stabilizer code associated with stabilizer group
given by 
\[
\S = \{ U_{L\a} V_{M \a } : \a \in \GF[n]{2} \},
\]
where $M = S L + J$. The corresponding 1-forbidden set is given by 
\[
F_1 = \{ a M^T e_i + b L^T e_i : a,b \in \GF{2} \}.
\]

It can be easily verified that $F_{d + 1} = F_d + F_1$. Now, let $W_d
= \{ wt(\a) : \a \in F_d \}$. It can be easily checked that $W_1 =
\{1,2,3,n-3,n-2,n-1\}$ and $W_2 = \{ 1,2, 3, 4, 5, 6, n-6,n-5,n- 4,
n-3 , n-2 , n-1 \}$.

If $B$ is a subset of $\GF[n]{2}$ such that for $u \in B - B$ we have
$wt(u) \not \in W_2$ then $B$ is the Fourier description of a 1-error
correcting quantum code.  A natural approach to finding large Fourier
descriptions $B$ is to solve the following combinatorial problem.

\begin{problem}\label{prob1}
Construct a family of subsets $\mathfrak{F}$ of
$\{1,2,\ldots,n\}$ such that for all $S_1,S_2 \in \mathfrak{F}$, $S_1
\ne S_2$ we have
\[ 
\# (S_1 \setminus S_2) + \# (S_2\setminus S_1) \not\in W_2. 
\]
\end{problem}

Given such a collection of subsets $\mathfrak{F}$, it is clear that
the set $B$ defined as
\begin{eqnarray}\label{B-sets}
 B =\{ \sum_{i \in S} e_i : S \in \mathfrak{F}\}
\end{eqnarray}
will yield the Fourier description of a 1-error correcting quantum
code by Theorem~\ref{distdtheorem}. For, the condition on the family
of subsets $\mathfrak{F}$ will ensure that the weight of any element
in $B - B$ does not lie in the set $\{
1,2,3,4,5,6,n-6,n-5,n-4,n-3,n-2,n-1\}$ and hence $(B - B)\cap
F_2=\emptyset$.

As our first example we describe a $((15,8,3))$ code. For $n=15$ it
suffices to construct a family of $8$ subsets $\mathfrak{F}$ such that
for any two distinct subsets $S_1,S_2\in\mathfrak{F}$ we have $\# (S_1
\setminus S_2) + \# (S_2\setminus S_1)\in\{7,8\}$. Then, $B$ defined
by Equation~\ref{B-sets} will be the Fourier description of a
$((15,8,3))$ code. The eight subsets of $\{1,2,\ldots,15\}$ that we
pick are as follows:

\begin{enumerate}
\item[] $S_1 = \{1,2,3,4,13\}$,  $S_2 = \{5,6,7,8,13\}$,  
$S_3 = \{9,10,11,12,13\}$,
\item[] $S_4 = \{1,2,5,6,9,10\}$, $S_5 = \{1,2,7,8,11,12\}$, $S_6 =
\{3,4,7,8,9,10\}$,
\item[] $S_7 = \{3,4,5,6,11,12\}$,  $S_8 = \{14,15\}$.
\end{enumerate}

In order to construct such nonstabilizer codes for general $n$ we need
to construct explicit set families $\mathfrak{F}$ as a solution to
Problem~\ref{prob1}. To this we describe a general method and use it
to construct a $((33,155,3))$ code. More precisely, we will seek a
special solution of Problem~\ref{prob1} in which all the sets in
$\mathfrak{F}$ are of the same cardinality.


Consider the case $n = 33$. In our explicit construction we consider
only subsets of $\{ 1, 2, \ldots , 32 \}$. Consider the vector space
$\GF[5]{2}$. As sets, $\GF[5]{2}$ and $\{1,2,\ldots,32\}$ are of the
same size and can be identified using any 1-1 correspondence. Our goal
is essentially to find a family of subsets of $\GF[5]{2}$ satisfying
the above conditions.  Let $\mathfrak{F}$ be the family of all 3
dimensional subspaces of $\GF[5]{2}$. Since any $S \in \mathfrak{F}$
is a vector space over $\GF{2}$ of dimension three, we have $\# S =
2^3 = 8$.  Moreover any two distinct subspaces can have at most 4
vectors in common. Hence for every pair of distinct sets $ S_1,S_2 \in
\mathfrak{F}$ we have $\# S_1 \cap S_2 \leq 4$. Consequently, for
distinct sets $ S_1,S_2 \in \mathfrak{F}$ we have 
\[
8\leq \# (S_1 \setminus S_2) + \# (S_2\setminus S_1) \leq 14.
\]

Thus, for distinct sets $ S_1,S_2 \in \mathfrak{F}$ $\# (S_1 \setminus
S_2) + \# (S_2\setminus S_1)\not\in W_2$ and hence the corresponding
Fourier description $B$ gives rise to a 1-error correcting code. 

Now, to find the size of the set $B$ which is the dimension of the
code we have to find the size of $\mathfrak{F}$. The following general
theorem gives the exact size.

\begin{theorem} \label{gntheorem}
  Consider the vector space $\GF[m]{q}$. The number of subspace of dimension
$r$ is given by
\[
N(n,q,r) = \frac{(q^m - 1)(q^{m-1} -1 ) \ldots (q -1) }
            { (q^r -1)(q^{r-1} -1 ) \ldots (q-1) ( q^{m-r} -1 ) (q^{m-r-1} -1 )
              \ldots (q -1 ) } 
\]
\end{theorem}
\begin{proof}
  Let $\e_1,\e_2,\ldots,\e_m$ be the standard basis for $\GF[m]{q}$.
  Let $\mathfrak{T}$ be the family of all $r$ dimensional subspaces of
  $\GF[m]{q}$.  We want to find $\# \mathfrak{T}$. Let $R$ be the
  subspace of $\GF[m]{q}$ spanned by the vectors
  $\e_1,\e_2,\ldots,\e_r$. Consider the group $G = \GLnq$. $G$ acts on
  $\mathfrak{T}$ transitively and hence the orbit of $R$ under the $G$
  action is the whole of $\mathfrak{T}$. Hence the number of element in
  $\mathfrak{T}$ is given by
  \[
  \# \mathfrak{T} = \frac{\# G}{\# G_R}
  \]
  where $G_R$ is the subgroup of $G$ that leaves $R$ invariant.
  
  Any element of $G_R$ is of the form 
  \[
  \left(
    \begin{array}{c|c}
      A&*\\
      \hline
      0 & B
    \end{array}
  \right)
  \]
  where $A$ and $B$ are $r \times r $ and $(m-r) \times (m-r)$ nonsingular matrices respectively and
  $*$ is any $r \times ( m -r) $ matrix. If $g(m) = \# \GLnq$ then we have 
  \[
  \# G_R = g(r) g(m-r) q^{(m-r)r}.
  \]
  So the problem reduces to finding $g(m)$. Let $A$ be any matrix in
  $\GLnq$. The first column of $A$ can be any one of the nonzero
  vectors is $\GF[m]{q}$. The are $q^m -1$ nonzero vectors.  Having
  fixed the first column $\a_1$, we have $q^m - q$ choices for the
  second column $\a_2$. Similarly there are $q^m -q^2$ choices for the
  third column and so on. Therefore the number of elements in $\GLnq$
  is given by
\[
g(m) = \prod_{i=0}^{m-1} (q^m - q^i).
\]

Therefore the size of $\mathfrak{T}$ is given by
\begin{eqnarray}
  \nonumber
  \# \mathfrak{T} &=& \frac{g(m)}{g(r)g(m-r) q^{(m-r)r}}\\
    \nonumber
    &=& \frac{(q^m - 1)(q^{m-1} -1 ) \ldots (q -1) }
            { (q^r -1)(q^{r-1} -1 ) \ldots (q-1) ( q^{m-r} -1 ) (q^{m-r-1} -1 )
              \ldots (q -1 ) }  
\end{eqnarray}
\end{proof}

From Theorem~\ref{gntheorem} we have 
\[
\# \mathfrak{F} = N(5,2,3) = 155
\] and hence the set $B$ defined as
\[
B = \{ \sum_{i \in S} e_i : S \in \mathfrak{F} \}
\] gives a $((33,155,3))$ code.

\begin{remark}
We can actually obtain a $((31,155,3))$ code as follows: construct the
same family of 3-dimensional subspaces of $\GF[5]{2}$ which are 155 in
number. Now, drop the extra coordinate, which will still result in 155
distinct subsets such that the symmetric difference of any pair of
these has weight in the range 7 to 13. Thus, we have a $((31,155,3))$
nonstabilizer code. We can easily extend this puncturing argument to
other nonstabilizer codes.
\end{remark}

\section{Encoding circuits for a class of nonstabilizer codes}

In this section we discuss the encoding algorithm for the class of
non-stabilizer codes defined in Section~\ref{explicit} and the
asymptotically good codes of Corollary~\ref{asymp-good}. Recall that
given an Gottesman subgroup, a code can be specified by giving its
Fourier description. We fix our encoding space to be $L^2(\GF[n]{q})$.
Let \mbox{$C = \{ \a | \sum_i a_i = 0 \}$} and $C^\perp$ the set $\{
\b | \b^T \a = 0 \textrm{ for all } \a \in C\}$ We restrict attention
to maximal Gottesman subgroups of the form

\[
\S = \{s_{a,b} = \omega( \a^T D \a ) U_\a V_{L\a + \b} | \a \in C, \b
\in C^\perp \}
\] 

where $D$ is an upper triangular matrix and $L = D+D^T$. Consider a
code $\C_n$ with Fourier description $B \subseteq \S$.  Recall that
the dimension of the code is $\#B$. Due to the isomorphism $\S \cong C
\times C^\perp \hookrightarrow \GF[n]{q} \times \GF[n]{q}$ we have the
character group of $\S$ as

\[
 \chi_{s_{\c,\d}}(s_{\a,\b}) = \chi_{\c,\d}(\a,\b) = \omega(\a^T\c +
\b^T \c).
\]

For $u \in \S$ define $\S_u$ to be the abelian group $\{ \chi_u(s) s |
s \in \S\}$. It is easy to see that $\S_u$ is also a maximal Gottesman
subgroup. In this notation we have $\S_1 = \S$. Let
$\mathcal{C}_{\S_u}$ denote the stabilizer (one dimensional) code
corresponding to the Gottesman subgroup $\S_u$.  Let
$\{\ket{\varphi_u}\}$ denote a (singleton) orthonormal basis for
$\mathcal{C}_{\S_u}$ for each $u\in\S$.

\begin{theorem}\label{onbasis}
  The vectors $\ket{\varphi_u}, u \in B$ forms an orthonormal basis
  for the code with Fourier description $B$.
\end{theorem}

\begin{proof}
  The projection operators for the code with Fourier description $B$
  is given by 

\[
P = \frac{1}{\#S} \sum_{u \in B} \sum_{s \in \S} \chi_u(s) s = \sum_{u
  \in B} P_u
\] 

where $P_u = \frac{1}{\#\S} \sum_{s \in \S} \chi_u(s) s$. Note that
$P_u$ is nothing but the projection operator corresponding to the
stabilizer code $\mathcal{C}_{\S_u}$. To prove that $\{\ket{\varphi_u}
| u \in B\}$ forms an orthonormal basis for the code given by $B$ it
suffices to show that
  
\[
  P_u P_v = 
\left\{
\begin{array}{cc}
    0&\textrm{if } u \neq v, \\
    P_u&\textrm{otherwise,}
\end{array}
\right.
\]
  
which is an immediate consequence of the following:
  
\[
  P_u P_v = P_{\chi_u * \chi_v} = \left\{\begin{array}{cc}
      0&\textrm{if } u \neq v,\\
      P_u&\textrm{otherwise.}
    \end{array}
  \right. 
\]

It follows that $\{\ket{\varphi_u}\mid u \in B\}$ is an orthonormal
basis for $\C_n$.
\end{proof}

For $s_{\a,\b} \in \S$ instead of writing $\S_{s_{\a\b}}$ we will
write $\S_{\a,\b}$. Similarly $\ket{\varphi_{\a,\b}}$ will be used to
denote $\ket{\varphi_{s_{\a,\b}}}$. It is easy to see that

\[
\ket{\varphi_{\c,\d}} = \sqrt{\frac{1}{\# C}}\sum_{{\bf x} \in C}
\omega(({\bf x} +\d)^T D ({\bf x} + \d) ) \overline{\omega}({\bf x}^T
c) \ket{{\bf x} + d} .
\]

Since the code has dimension $\# B$, we will assume that the encoding
message space is a Hilbert space of dimension $\# B$ with basis are
indexed by elements of $B$ viz. $\{ \ket{\c,\d} | s_{\c,\d} \in B\}$.
To summarize we have the following observation.

\begin{proposition}
  For a code with Fourier description $B$ then the encoding procedure
  is given by linear map with the following property $\ket{\c,\d}
  \mapsto \ket{\varphi_{\c,\d}}$ for all $s_{\c,\d} \in B$.
\end{proposition}


We give encoding circuits for the codes over $\GF{q}$. We assume that
some basic gates over $L^2(\GF{q})$, which we will define in a moment,
are given as black boxes. It is to be noted that for a fixed $q$ these
gates are easily implementable using the standard set of gates. The
most basic of the gates which we require is the inverter gate defined
as follows

\[
\mathcal{I} \ket{a} = \ket{-a}; \ a \in \GF{q}.
\]

Let $C\!-\!U$ and $C\!-\!V$ stand for the unitary transformation
define as follows

\[
C\!-\!U \ket{a}\ket{b} = \ket{a}\ket{ a + b}.
\] 

and 

\[
C\!-\!V\ket{a}\ket{b} = \omega(ab) \ket{a} \ket{b}; \ \ a,b \in
\GF{q}.
\]

Note that for $q=2$ the gate $C\!-\!U$ is just the $C\!-\!NOT$ gate
and the gate $C\!-\!V$ is the $Z$ gate. We also assume that the
operator $CC\!-\!U$ defined as is available:

\[
CC\!-\!U \ket{a,b,c} = \ket{a,b,c + ab}.
\] 

For $\GF{2}$ it is the $CC\!-\!NOT$ gate. Note that $CC\!-\!V$ defined
as

\[
CC\!-\!V \ket{a,b,c} = \omega( c + a b) \ket{a,b,c}
\]

can be defined using $CC\!-\!U,\mathcal{I}$ and $C\!-\!V$. See
Figure~\ref{figbasicgates} for a pictorial description.

We need one more operator which is the Fourier transform operator over
$\GF{q}$ which we will denote by $F$. It is defined as follows

\[
F \ket{a} = \frac{1}{\sqrt{q}} \sum_{x \in \GF{q}} \omega(ax) \ket{x}.
\]

Note that if $q = 2$ then $F$ is nothing but the Hadamard operator.
Consider any gate $C-f$ of two arguments $a,b \in \GF{q}$ defined as

\[
C\!-\!f \ket{a}\ket{b} = \ket{a}\ket{f(a,b)}.
\]    

We extend it to an operator $C\!-\!f_n$ acting on two arguments $\a,\b
\in \GF[n]{q}$ in a natural way as follows (see Figure~\ref{figCfn})

\[
C\!-\!f_n \ket{\a}\ket{\b} = C\!-\!f_n \ket{a_1a_2\ldots a_n}
\ket{b_1b_2\ldots b_n} = \ket{\a} \ket{f(a_1,b_1) , f(a_2,b_2), \ldots
,f(a_n,b_n)}.
\]

In a similar fashion we extend the gate $CC\!-\!f$ acting on $a,b,c
\in \GF{q}$ to $CC\!-\!f_n$ acting on $\a,\b,\c \in \GF[n]{q}$. In the
circuits we draw a $C\!-\!f_n$ gate with thick wires to indicate that
it takes a tuple from $\GF[n]{q}$.

From these gates we can construct a circuit that computes for any
$\a,\b \in \GF[n]{q}$ the dot product $\sum_i a_i b_i$ (see
Figure~\ref{figdotproduct}). The second circuit in the figure is a
symbolic representation of the circuit (note the thick lines). Using
this inner product circuit we can also define the circuit that takes a
$n\times n$ matrix $D$ and a vector $\a\in \GF[n]{q}$ and computes the
vector $D\a$ (see Figure~\ref{figmatvect} note the thick lines and the
cut).

We also need a circuit which will take the vector $\ket{0^n}$ and
generate the uniform superposition $\frac{1}{\sqrt{q^{n-1}}} \sum_{x
  \in C} \ket{x}$. This circuit is given in Figure~\ref{figC}.

Given the message $\c,\d$, we can describe the main steps of the
encoding algorithm as follows.

\begin{enumerate}
\item Initialize $\ket{R} := \ket{0^n}$
\item Apply $C$ on $R$ so that
  
\[
  \ket{R} = \sqrt{\frac{1}{q^{n-1}}}\sum_{\x \in C}\ket{\x}.
\]
  
\item Apply $C\!-\!V_n$ on $\ket{c}\ket{R}$ so that
    
\[
\ket{c}\ket{R} \mapsto \ket{c} \otimes (\frac{1}{\sqrt{q^{n-1}}}
\sum_{\x \in C} \omega(c^T x ) \ket{x}).
\]
  
\item Apply $C\!-\!U_n$ on $\ket{\d}\ket{R}$
    
\[
\ket{\d}\ket{R} \mapsto \ket{d} \otimes ( \frac{1}{\sqrt{q^{n-1}}}
    \sum_{\x \in C}\omega{\c^T x}\ket{\x + \d}.
\]
  
\item Apply the circuit in Figure~\ref{figmatvect} on
    $\ket{R}\ket{D}\ket{0^n}$ 
\[
\ket{R}\ket{D}\ket{0^n} \mapsto \frac{1}{\sqrt{q^{n-1}}}\sum_{\x \in
  C} \omega(\c^T \x) \ket{\x+\d,D(\x+\d)}\ket{D}.
\]
  
\item Apply $C\!-\!V_n$ on
    $\ket{R_1}=\frac{1}{\sqrt{q^{n-1}}}\sum_{\x \in C} \omega(\c^T \x)
    \ket{\x+\d,D(\x+\d)}$ to get
    
\[
\ket{R_1} \mapsto \frac{1}{\sqrt{q^{n-1}}}\sum_{\x \in C}
\omega((\x+\d)^TD(\x+\d) + \c^T \x) \ket{\x+\d,D(\x+\d)}.
\]
\end{enumerate}

\section{Decoding for a class of nonstabilizer codes}

Let $\C\subseteq\Hin$ be a $t$-error correcting quantum code
satisfying the conditions of Theorem~\ref{distdtheorem}, with Fourier
description $B$, and such that its projection $P$ has support in the
Gottesman subgroup $\S$ of the error group $\E$. The two conditions that
$\C$ satisfies are:

\begin{enumerate}
  
\item 
For each $g \in \S$ such that $\wt(g)\leq 2t$
\[
\chi_g( u_1^{-1} u_2) = 1 \textrm{ for all } u_1,u_2 \in B. 
\]

\item 
For each $g \in \E \setminus \S$ such that $\wt(g)\leq 2t$, and for 
$s\in S$ we have
\[
\sum_{u_1,u_2 \in B} \chi_{u_2}(s) \sum_{s_1 }\gamma(g,s_1)
\chi_{s_1} (u_1^{-1} u_2 ) = 0,
\]
where $\gamma(g,s)$ is the scalar such that $\gamma(g,s) s g = gs$.
\end{enumerate}

For $u \in \S$ let $\S_u$ be the Gottesman group $\{ \chi_u(s) s | s \in
\S\}$. 

The projection operator for the code $\C_n$

\[
P = \frac{1}{\#\S} \sum_{u \in B} \sum_{s \in \S} \chi_u(s) s = \sum_{u
  \in B} P_u,
\] 

where $P_u = \frac{1}{\#\S} \sum_{s \in \S} \chi_u(s) s$. Note that
$P_u$ is the projection operator corresponding to the stabilizer code
$\C_u$ with $\S_u$ as stabilizer group for each $u\in B$. As argued in
Theorem~\ref{onbasis}, $P_uP_v=0$ for $u\neq v\in B$. Let $D_u$ denote
an orthonormal basis for $\C_u$, $u\in B$. Then $D=\bigcup_{u\in B}
D_u$ is an orthonormal basis for $\C$.

It suffices to describe the decoding procedure for the encoded message
as a basis element from $D$ and error $g\in\E$ such that $\wt(g)\leq
t$. W.l.o.g.\ let $\ket{\psi}\in D_u$ for some $u\in B$ be the encoded
message, and let $g\in\E$ of weight at most $t$ be the error operator.
The decoding procedure takes $g\ket{\psi}=\ket{\psi'}$ as input and
outputs $\ket{\psi}$. Let $\{s_1,s_2,\ldots,s_k\}$ be an independent
generator set for $\S$. Notice that for $1\leq i\leq k$

\[
s_ig\ket{\psi}=\overline{\gamma(s_i,g)}\chi_u(s_i)g\ket{\psi}.
\]

Thus, $g\ket{\psi}$ is an eigen vector for operator $s_i$ with eigen
value $\overline{\gamma(s_i,g)}\chi_u(s_i)$.

The decoding procedure will carry out the following steps. It uses as
subroutine the phase estimation algorithm of Kitaev (c.f
\cite{NC99}).

\begin{enumerate}
\item Let $\ket{\psi'}$ be the received state.
  
\item Apply $s_i$ successively, for each $1\leq i\leq k$, and when
  $s_i$ is applied run Kitaev's phase estimation algorithm to compute
  the eigen value $\alpha_i=\overline{\gamma(s_i,g)}\chi_u(s_i)$. Let
  $\ket{\rho}$ be the resulting state.
  
\item If $g=I$ and some $u\in B$ constitute a solution to the system
  of $k$ group equations (using a classical algorithm that searches
  through $B$):
\[
\alpha_i=\overline{\gamma(s_i,g)}\chi_u(s_i), 1\leq i\leq k,
\]

then apply $s_i^{-1}$ for each $1\leq i\leq k$ to the state
$\ket{\rho}$ and output that as the decoded state.

\item If $g=I$ does not give a solution to the $k$ equations, find
  (using a classical algorithm that searches through $B$) a $g\neq
  I\in\E$ and the corresponding unique $u\in B$ which are a solution
  to the $k$ equations.  Apply $g^{-1}$ to the current state. Then
  apply $s_i^{-1}$ for each $1\leq i\leq k$, and output that as the
  decoded state.
\end{enumerate}

We now argue the correctness of the procedure. Firstly, notice that if
the error operator is $g\in\S$ such that $\wt(g)\leq t$, then by
assumption $\chi_g(u_2) = \chi_g(u_1)$  for all $u_1,u_2 \in
B$. Denote this scalar by $\lambda$. Notice that for any state
$\ket{\varphi}\in\C$, $g\ket{\varphi}=\lambda\ket{\varphi}$. Which
means that $g$ introduces only an overall phase. We establish the
following claim from which the correctness of the procedure follows.

\begin{claim}\label{decodeproof}
\begin{enumerate}
\item If $g=I$ and $u\in B$ is a solution to the $k$ group equations
  given above, then $u\in B$ is the unique solution, and there is no
  $g\not\in\S$ which is a solution to the equations.
\item If $g=I$ is not part of a solution to the $k$ equations, then
  there is a unique $u\in B$ and some $g\not\in\S$ that form a
  solution such that $\wt(g)\leq t$.
\end{enumerate}
\end{claim}

\begin{claimproof}
  For the first part, assume that $g=I$ and $u\in B$ is a solution to
  the $k$ group equations, and some $g\not\in\S$ and $u'\in B$ is
  another solution. Then we have
  $\gamma(g,s_i)\chi_u(s_i)\overline{\chi_u'(s_i)}=1, 1\leq i\leq k$.
  Since $s_1,s_2,\ldots,s_k$ generate $\S$, it implies
  $\gamma(g,s_i)\chi_u(s)\overline{\chi_u'(s)}=1$ for all $s\in\S$. It
  is easy to see that this contradict the second condition of
  Theorem~\ref{distdtheorem} for the element $g\not\in\S$ of
  $\wt(g)\leq t$.

For the second part, notice that any solution $g$ of $\wt(g)\leq t$ to
the equations is not in $\S$. For, if $g'\in\S$ and $u\in B$ are a
solution then so is $g=I$ and $u\in B$, because $\gamma(g',s)=1$ for
all $s\in\S$. Assume to the contrary that there are two distinct
solutions $g_1\not\in\S$ and $u\in B$ and $g_2\not\in\S$ and $u'\in
B$, where $\wt(g_1)\leq t$ and $\wt(g_2)\leq t$. Then, as before, the
$k$ equations will yield

\[
\gamma(g_1,s)\chi_u(s)=\gamma(g_2,s)\chi_u'(s), ~~~~\forall s\in\S.
\]

By rearranging terms we get $\gamma(g_1^{-1}g_2,s)\chi_s(u'u^{-1})=1$
for all $s\in\S$. 

Now, if $g_1^{-1}g_2\not\in\S$, then this again contradicts the second
condition of Theorem~\ref{distdtheorem} for the element
$g_1^{-1}g_2\not\in\S$ of $\wt(g_1^{-1}g_2)\leq 2t$.

Next, suppose that $g_1^{-1}g_2\in\S$. Then we get $\chi_s(u'u^{-1}=1$
for all $s\in\S$ implying that $u=u'$. Thus $u$ is unique. Notice that
by the first condition of Theorem~\ref{distdtheorem}, since
$\wt(g_1^{-1}g_2)\leq 2t$, the effect of applying the error
$g_1^{-1}g_2$ to a state in $\C$ only introduces an overall phase. Thus,
decoding with either $g_1^{-1}$ or $g_2^{-1}$ will coincide upto an
overall phase. This completes the proof of the claim and correctness
of the decoding procedure. 
\end{claimproof}

To analyze the efficiency of the decoding procedure, we recall from
\cite{NC99} that the phase estimation quantum circuit is efficient
(polynomial size in $n$). However, solving the $k$ group equations
involves exhaustive enumeration. This takes time $O(n^{O(d)}.  \# B)$,
which is also the dominant term in the entire time bound.

\bibliographystyle{alpha}
\bibliography{NonStab}

\newpage

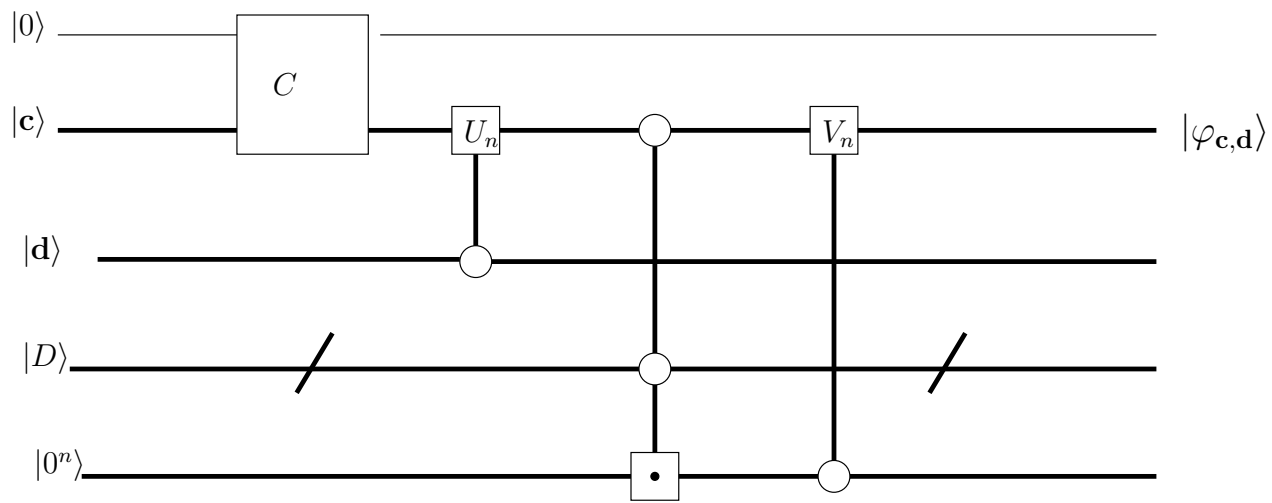
\begin{figure}[h]
  \input{encode.pstex_t}
  \caption{Complete encoding circuit}
  \label{figencode}
\end{figure}

See Figure~\ref{figencode} for a complete circuit. The other figures
are the building blocks. Note that the extra$\ket{D(\x+\d)}$ can be
removed by inverting it and then applying the circuit in
fig~\ref{figmatvect} on $\ket{\x+\d,D(\x+\d)}\ket{D}$.\\

\vspace{1cm}

\begin{figure}[h]
  \input{basicgates.pstex_t}
  \caption{Basic Gates}
  \label{figbasicgates}
\end{figure}

\begin{figure}[h]
  \input{Cfn.pstex_t}
  \caption{Circuit computing $C\!-\!f_n$ from $C\!-\!f$}
  \label{figCfn}
\end{figure}

\begin{figure}[h]
  \input{dotproduct.pstex_t}
  \caption{Circuit computing the dot product $\a^T\b$}
  \label{figdotproduct}
\end{figure}

\begin{figure}[h]
  \input{matvect.pstex_t}  
  \caption{Circuit to compute $D\x$}
  \label{figmatvect}
\end{figure}

\begin{figure}[h]
  \input{uniformC.pstex_t}
  \caption{Circuit generating $\sqrt{\frac{1}{q^{n-1}}} \sum_{x \in C} \ket{x}$}
  \label{figuniformC}
\end{figure}

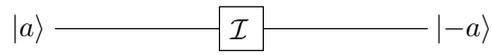
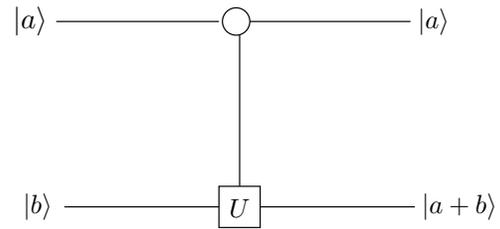
\end{document}

%% file: basicgates.pstex_t
\begin{picture}(0,0)%
\includegraphics{basicgates.pstex}%
\end{picture}%
\setlength{\unitlength}{3947sp}%
\begingroup\makeatletter\ifx\SetFigFont\undefined
\def\x#1#2#3#4#5#6#7\relax{\def\x{#1#2#3#4#5#6}}%
\expandafter\x\fmtname xxxxxx\relax \def\y{splain}%
\ifx\x\y   
\gdef\SetFigFont#1#2#3{%
  \ifnum #1<17\tiny\else \ifnum #1<20\small\else
  \ifnum #1<24\normalsize\else \ifnum #1<29\large\else
  \ifnum #1<34\Large\else \ifnum #1<41\LARGE\else
     \huge\fi\fi\fi\fi\fi\fi
  \csname #3\endcsname}%
\else
\gdef\SetFigFont#1#2#3{\begingroup
  \count@#1\relax \ifnum 25<\count@\count@25\fi
  \def\x{\endgroup\@setsize\SetFigFont{#2pt}}%
  \expandafter\x
    \csname \romannumeral\the\count@ pt\expandafter\endcsname
    \csname @\romannumeral\the\count@ pt\endcsname
  \csname #3\endcsname}%
\fi
\fi\endgroup
\begin{picture}(6922,4971)(1126,-6569)
\put(6735,-4792){\makebox(0,0)[lb]{\smash{\SetFigFont{11}{13.2}{rm}{\color[rgb]{0,0,0}$F$}%
}}}
\put(5326,-4769){\makebox(0,0)[lb]{\smash{\SetFigFont{11}{13.2}{rm}{\color[rgb]{0,0,0}$\ket{a}$}%
}}}
\put(8048,-4769){\makebox(0,0)[lb]{\smash{\SetFigFont{11}{13.2}{rm}{\color[rgb]{0,0,0}$F\ket{a}$}%
}}}
\put(2486,-5546){\makebox(0,0)[lb]{\smash{\SetFigFont{10}{12.0}{rm}{\color[rgb]{0,0,0}$V$}%
}}}
\put(1191,-5525){\makebox(0,0)[lb]{\smash{\SetFigFont{10}{12.0}{rm}{\color[rgb]{0,0,0}$\ket{b}$}%
}}}
\put(1126,-4359){\makebox(0,0)[lb]{\smash{\SetFigFont{11}{13.2}{rm}{\color[rgb]{0,0,0}$\ket{a}$}%
}}}
\put(3695,-5525){\makebox(0,0)[lb]{\smash{\SetFigFont{10}{12.0}{rm}{\color[rgb]{0,0,0}$\omega(ab)\ket{b}$}%
}}}
\put(3669,-4359){\makebox(0,0)[lb]{\smash{\SetFigFont{10}{12.0}{rm}{\color[rgb]{0,0,0}$\ket{a}$}%
}}}
\put(6686,-2921){\makebox(0,0)[lb]{\smash{\SetFigFont{10}{12.0}{rm}{\color[rgb]{0,0,0}$U$}%
}}}
\put(5391,-2900){\makebox(0,0)[lb]{\smash{\SetFigFont{10}{12.0}{rm}{\color[rgb]{0,0,0}$\ket{b}$}%
}}}
\put(5326,-1734){\makebox(0,0)[lb]{\smash{\SetFigFont{11}{13.2}{rm}{\color[rgb]{0,0,0}$\ket{a}$}%
}}}
\put(7895,-2900){\makebox(0,0)[lb]{\smash{\SetFigFont{10}{12.0}{rm}{\color[rgb]{0,0,0}$\ket{a+b}$}%
}}}
\put(7869,-1734){\makebox(0,0)[lb]{\smash{\SetFigFont{10}{12.0}{rm}{\color[rgb]{0,0,0}$\ket{a}$}%
}}}
\put(2504,-1842){\makebox(0,0)[lb]{\smash{\SetFigFont{11}{13.2}{rm}{\color[rgb]{0,0,0}$\mathcal{I}$}%
}}}
\put(1126,-1819){\makebox(0,0)[lb]{\smash{\SetFigFont{11}{13.2}{rm}{\color[rgb]{0,0,0}$\ket{a}$}%
}}}
\put(3789,-1819){\makebox(0,0)[lb]{\smash{\SetFigFont{11}{13.2}{rm}{\color[rgb]{0,0,0}$\ket{-a}$}%
}}}
\put(6251,-6511){\makebox(0,0)[lb]{\smash{\SetFigFont{12}{14.4}{rm}{\color[rgb]{0,0,0}Fourier Transform gate}%
}}}
\put(6401,-3561){\makebox(0,0)[lb]{\smash{\SetFigFont{12}{14.4}{rm}{\color[rgb]{0,0,0}$C\!-U\!$ gate}%
}}}
\put(2101,-3561){\makebox(0,0)[lb]{\smash{\SetFigFont{12}{14.4}{rm}{\color[rgb]{0,0,0}Invertor gate}%
}}}
\put(2101,-6511){\makebox(0,0)[lb]{\smash{\SetFigFont{12}{14.4}{rm}{\color[rgb]{0,0,0}$C\!-\!V$ gate}%
}}}
\end{picture}

%% file: Cfn.pstex_t
\begin{picture}(0,0)%
\includegraphics{Cfn.pstex}%
\end{picture}%
\setlength{\unitlength}{3947sp}%
\begingroup\makeatletter\ifx\SetFigFont\undefined
\def\x#1#2#3#4#5#6#7\relax{\def\x{#1#2#3#4#5#6}}%
\expandafter\x\fmtname xxxxxx\relax \def\y{splain}%
\ifx\x\y   
\gdef\SetFigFont#1#2#3{%
  \ifnum #1<17\tiny\else \ifnum #1<20\small\else
  \ifnum #1<24\normalsize\else \ifnum #1<29\large\else
  \ifnum #1<34\Large\else \ifnum #1<41\LARGE\else
     \huge\fi\fi\fi\fi\fi\fi
  \csname #3\endcsname}%
\else
\gdef\SetFigFont#1#2#3{\begingroup
  \count@#1\relax \ifnum 25<\count@\count@25\fi
  \def\x{\endgroup\@setsize\SetFigFont{#2pt}}%
  \expandafter\x
    \csname \romannumeral\the\count@ pt\expandafter\endcsname
    \csname @\romannumeral\the\count@ pt\endcsname
  \csname #3\endcsname}%
\fi
\fi\endgroup
\begin{picture}(5202,8167)(1501,-8698)
\put(4051,-8611){\makebox(0,0)[lb]{\smash{\SetFigFont{12}{14.4}{rm}{\color[rgb]{0,0,0}$f_n$}%
}}}
\put(6076,-7186){\makebox(0,0)[lb]{\smash{\SetFigFont{12}{14.4}{rm}{\color[rgb]{0,0,0}$\ket{\a}$}%
}}}
\put(6076,-7861){\makebox(0,0)[lb]{\smash{\SetFigFont{12}{14.4}{rm}{\color[rgb]{0,0,0}$\ket{\b}$}%
}}}
\put(6151,-8611){\makebox(0,0)[lb]{\smash{\SetFigFont{12}{14.4}{rm}{\color[rgb]{0,0,0}$\ket{\c + f_n(\a,\b)}$}%
}}}
\put(1876,-7111){\makebox(0,0)[lb]{\smash{\SetFigFont{12}{14.4}{rm}{\color[rgb]{0,0,0}$\ket{\a}$}%
}}}
\put(1876,-7786){\makebox(0,0)[lb]{\smash{\SetFigFont{12}{14.4}{rm}{\color[rgb]{0,0,0}$\ket{\b}$}%
}}}
\put(1876,-8536){\makebox(0,0)[lb]{\smash{\SetFigFont{12}{14.4}{rm}{\color[rgb]{0,0,0}$\ket{\c}$}%
}}}
\put(4369,-5380){\makebox(0,0)[lb]{\smash{\SetFigFont{9}{10.8}{rm}{\color[rgb]{0,0,0}$f$}%
}}}
\put(1501,-5308){\makebox(0,0)[lb]{\smash{\SetFigFont{9}{10.8}{rm}{\color[rgb]{0,0,0}$\ket{c_2}$}%
}}}
\put(6703,-5360){\makebox(0,0)[lb]{\smash{\SetFigFont{12}{14.4}{rm}{\color[rgb]{0,0,0}$\ket{c_2 + f(a_2,b_2)}$}%
}}}
\put(6635,-3078){\makebox(0,0)[lb]{\smash{\SetFigFont{9}{10.8}{rm}{\color[rgb]{0,0,0}$\ket{b_1}$}%
}}}
\put(1501,-3026){\makebox(0,0)[lb]{\smash{\SetFigFont{9}{10.8}{rm}{\color[rgb]{0,0,0}$\ket{b_2}$}%
}}}
\put(6650,-2610){\makebox(0,0)[lb]{\smash{\SetFigFont{10}{12.0}{rm}{\color[rgb]{0,0,0}$\ket{b_1}$}%
}}}
\put(1501,-2557){\makebox(0,0)[lb]{\smash{\SetFigFont{9}{10.8}{rm}{\color[rgb]{0,0,0}$\ket{b_1}$}%
}}}
\put(6670,-2141){\makebox(0,0)[lb]{\smash{\SetFigFont{10}{12.0}{rm}{\color[rgb]{0,0,0}$\ket{a_n}$}%
}}}
\put(1501,-2093){\makebox(0,0)[lb]{\smash{\SetFigFont{9}{10.8}{rm}{\color[rgb]{0,0,0}$\ket{a_n}$}%
}}}
\put(6650,-1146){\makebox(0,0)[lb]{\smash{\SetFigFont{10}{12.0}{rm}{\color[rgb]{0,0,0}$\ket{a_2}$}%
}}}
\put(1501,-1107){\makebox(0,0)[lb]{\smash{\SetFigFont{9}{10.8}{rm}{\color[rgb]{0,0,0}$\ket{a_2}$}%
}}}
\put(6650,-678){\makebox(0,0)[lb]{\smash{\SetFigFont{10}{12.0}{rm}{\color[rgb]{0,0,0}$\ket{a_1}$}%
}}}
\put(1501,-639){\makebox(0,0)[lb]{\smash{\SetFigFont{9}{10.8}{rm}{\color[rgb]{0,0,0}$\ket{a_1}$}%
}}}
\put(6703,-6297){\makebox(0,0)[lb]{\smash{\SetFigFont{12}{14.4}{rm}{\color[rgb]{0,0,0}$\ket{c_n + f(a_n,b_n)}$}%
}}}
\put(5423,-6316){\makebox(0,0)[lb]{\smash{\SetFigFont{9}{10.8}{rm}{\color[rgb]{0,0,0}$f$}%
}}}
\put(1501,-6245){\makebox(0,0)[lb]{\smash{\SetFigFont{9}{10.8}{rm}{\color[rgb]{0,0,0}$\ket{c_n}$}%
}}}
\put(3433,-4736){\makebox(0,0)[lb]{\smash{\SetFigFont{9}{10.8}{rm}{\color[rgb]{0,0,0}$f$}%
}}}
\put(6703,-4717){\makebox(0,0)[lb]{\smash{\SetFigFont{12}{14.4}{rm}{\color[rgb]{0,0,0}$\ket{c_1 + f(a_1,b_1)}$}%
}}}
\put(1501,-4664){\makebox(0,0)[lb]{\smash{\SetFigFont{9}{10.8}{rm}{\color[rgb]{0,0,0}$\ket{c_1}$}%
}}}
\put(6668,-4133){\makebox(0,0)[lb]{\smash{\SetFigFont{10}{12.0}{rm}{\color[rgb]{0,0,0}$\ket{b_n}$}%
}}}
\put(1501,-4079){\makebox(0,0)[lb]{\smash{\SetFigFont{9}{10.8}{rm}{\color[rgb]{0,0,0}$\ket{b_n}$}%
}}}
\end{picture}

%% file: dotproduct.pstex_t
\begin{picture}(0,0)%
\includegraphics{dotproduct.pstex}%
\end{picture}%
\setlength{\unitlength}{3947sp}%
\begingroup\makeatletter\ifx\SetFigFont\undefined
\def\x#1#2#3#4#5#6#7\relax{\def\x{#1#2#3#4#5#6}}%
\expandafter\x\fmtname xxxxxx\relax \def\y{splain}%
\ifx\x\y   
\gdef\SetFigFont#1#2#3{%
  \ifnum #1<17\tiny\else \ifnum #1<20\small\else
  \ifnum #1<24\normalsize\else \ifnum #1<29\large\else
  \ifnum #1<34\Large\else \ifnum #1<41\LARGE\else
     \huge\fi\fi\fi\fi\fi\fi
  \csname #3\endcsname}%
\else
\gdef\SetFigFont#1#2#3{\begingroup
  \count@#1\relax \ifnum 25<\count@\count@25\fi
  \def\x{\endgroup\@setsize\SetFigFont{#2pt}}%
  \expandafter\x
    \csname \romannumeral\the\count@ pt\expandafter\endcsname
    \csname @\romannumeral\the\count@ pt\endcsname
  \csname #3\endcsname}%
\fi
\fi\endgroup
\begin{picture}(6665,8193)(226,-7723)
\put(3901,-4936){\makebox(0,0)[lb]{\smash{\SetFigFont{12}{14.4}{rm}{\color[rgb]{0,0,0}$U$}%
}}}
\put(6846,-4161){\makebox(0,0)[lb]{\smash{\SetFigFont{12}{14.4}{rm}{\color[rgb]{0,0,0}$\ket{b_n}$}%
}}}
\put(6891,-4911){\makebox(0,0)[lb]{\smash{\SetFigFont{14}{16.8}{rm}{\color[rgb]{0,0,0}$\ket{c + \a^T\b}$}%
}}}
\put(6803,-2811){\makebox(0,0)[lb]{\smash{\SetFigFont{12}{14.4}{rm}{\color[rgb]{0,0,0}$\ket{b_1}$}%
}}}
\put(6825,-2211){\makebox(0,0)[lb]{\smash{\SetFigFont{12}{14.4}{rm}{\color[rgb]{0,0,0}$\ket{b_1}$}%
}}}
\put(6849,-1611){\makebox(0,0)[lb]{\smash{\SetFigFont{12}{14.4}{rm}{\color[rgb]{0,0,0}$\ket{a_n}$}%
}}}
\put(6824,-336){\makebox(0,0)[lb]{\smash{\SetFigFont{12}{14.4}{rm}{\color[rgb]{0,0,0}$\ket{a_2}$}%
}}}
\put(6824,264){\makebox(0,0)[lb]{\smash{\SetFigFont{12}{14.4}{rm}{\color[rgb]{0,0,0}$\ket{a_1}$}%
}}}
\put(2701,-4936){\makebox(0,0)[lb]{\smash{\SetFigFont{12}{14.4}{rm}{\color[rgb]{0,0,0}$U$}%
}}}
\put(5251,-4936){\makebox(0,0)[lb]{\smash{\SetFigFont{12}{14.4}{rm}{\color[rgb]{0,0,0}$U$}%
}}}
\put(1726,-6136){\makebox(0,0)[lb]{\smash{\SetFigFont{12}{14.4}{rm}{\color[rgb]{0,0,0}$\ket{\a}$}%
}}}
\put(1726,-6811){\makebox(0,0)[lb]{\smash{\SetFigFont{12}{14.4}{rm}{\color[rgb]{0,0,0}$\ket{\b}$}%
}}}
\put(1726,-7561){\makebox(0,0)[lb]{\smash{\SetFigFont{12}{14.4}{rm}{\color[rgb]{0,0,0}$\ket{c}$}%
}}}
\put(5926,-6211){\makebox(0,0)[lb]{\smash{\SetFigFont{12}{14.4}{rm}{\color[rgb]{0,0,0}$\ket{\a}$}%
}}}
\put(5926,-6886){\makebox(0,0)[lb]{\smash{\SetFigFont{12}{14.4}{rm}{\color[rgb]{0,0,0}$\ket{\b}$}%
}}}
\put(6001,-7636){\makebox(0,0)[lb]{\smash{\SetFigFont{12}{14.4}{rm}{\color[rgb]{0,0,0}$\ket{c + \a^T\b}$}%
}}}
\put(226,-4094){\makebox(0,0)[lb]{\smash{\SetFigFont{12}{14.4}{rm}{\color[rgb]{0,0,0}$\ket{b_n}$}%
}}}
\put(226,-2744){\makebox(0,0)[lb]{\smash{\SetFigFont{12}{14.4}{rm}{\color[rgb]{0,0,0}$\ket{b_2}$}%
}}}
\put(226,-2144){\makebox(0,0)[lb]{\smash{\SetFigFont{12}{14.4}{rm}{\color[rgb]{0,0,0}$\ket{b_1}$}%
}}}
\put(226,-1547){\makebox(0,0)[lb]{\smash{\SetFigFont{12}{14.4}{rm}{\color[rgb]{0,0,0}$\ket{a_n}$}%
}}}
\put(226,-286){\makebox(0,0)[lb]{\smash{\SetFigFont{12}{14.4}{rm}{\color[rgb]{0,0,0}$\ket{a_2}$}%
}}}
\put(226,314){\makebox(0,0)[lb]{\smash{\SetFigFont{12}{14.4}{rm}{\color[rgb]{0,0,0}$\ket{a_1}$}%
}}}
\put(226,-4844){\makebox(0,0)[lb]{\smash{\SetFigFont{12}{14.4}{rm}{\color[rgb]{0,0,0}$\ket{c}$}%
}}}
\end{picture}

%% file: matvect.pstex_t
\begin{picture}(0,0)%
\includegraphics{matvect.pstex}%
\end{picture}%
\setlength{\unitlength}{3947sp}%
\begingroup\makeatletter\ifx\SetFigFont\undefined
\def\x#1#2#3#4#5#6#7\relax{\def\x{#1#2#3#4#5#6}}%
\expandafter\x\fmtname xxxxxx\relax \def\y{splain}%
\ifx\x\y   
\gdef\SetFigFont#1#2#3{%
  \ifnum #1<17\tiny\else \ifnum #1<20\small\else
  \ifnum #1<24\normalsize\else \ifnum #1<29\large\else
  \ifnum #1<34\Large\else \ifnum #1<41\LARGE\else
     \huge\fi\fi\fi\fi\fi\fi
  \csname #3\endcsname}%
\else
\gdef\SetFigFont#1#2#3{\begingroup
  \count@#1\relax \ifnum 25<\count@\count@25\fi
  \def\x{\endgroup\@setsize\SetFigFont{#2pt}}%
  \expandafter\x
    \csname \romannumeral\the\count@ pt\expandafter\endcsname
    \csname @\romannumeral\the\count@ pt\endcsname
  \csname #3\endcsname}%
\fi
\fi\endgroup
\begin{picture}(6675,8401)(601,-8023)
\put(5851,-7186){\makebox(0,0)[lb]{\smash{\SetFigFont{12}{14.4}{rm}{\color[rgb]{0,0,0}$\ket{D}$}%
}}}
\put(5926,-7936){\makebox(0,0)[lb]{\smash{\SetFigFont{12}{14.4}{rm}{\color[rgb]{0,0,0}$\ket{\c + D\a}$}%
}}}
\put(1651,-7861){\makebox(0,0)[lb]{\smash{\SetFigFont{12}{14.4}{rm}{\color[rgb]{0,0,0}$\ket{\c}$}%
}}}
\put(5851,-6511){\makebox(0,0)[lb]{\smash{\SetFigFont{12}{14.4}{rm}{\color[rgb]{0,0,0}$\ket{\a}$}%
}}}
\put(1651,-6436){\makebox(0,0)[lb]{\smash{\SetFigFont{12}{14.4}{rm}{\color[rgb]{0,0,0}$\ket{\a}$}%
}}}
\put(1651,-7111){\makebox(0,0)[lb]{\smash{\SetFigFont{12}{14.4}{rm}{\color[rgb]{0,0,0}$\ket{D}$}%
}}}
\put(601,-528){\makebox(0,0)[lb]{\smash{\SetFigFont{12}{14.4}{rm}{\color[rgb]{0,0,0}$\ket{{\bf d_1}}$}%
}}}
\put(601,-2478){\makebox(0,0)[lb]{\smash{\SetFigFont{12}{14.4}{rm}{\color[rgb]{0,0,0}$\ket{{\bf d_n}}$}%
}}}
\put(601,222){\makebox(0,0)[lb]{\smash{\SetFigFont{12}{14.4}{rm}{\color[rgb]{0,0,0}$\ket{\a}$}%
}}}
\put(7276,164){\makebox(0,0)[lb]{\smash{\SetFigFont{14}{16.8}{rm}{\color[rgb]{0,0,0}$\ket{\a}$}%
}}}
\put(601,-3075){\makebox(0,0)[lb]{\smash{\SetFigFont{12}{14.4}{rm}{\color[rgb]{0,0,0}$\ket{c_1}$}%
}}}
\put(601,-3661){\makebox(0,0)[lb]{\smash{\SetFigFont{12}{14.4}{rm}{\color[rgb]{0,0,0}$\ket{c_2}$}%
}}}
\put(626,-1203){\makebox(0,0)[lb]{\smash{\SetFigFont{12}{14.4}{rm}{\color[rgb]{0,0,0}$\ket{{\bf d_2}}$}%
}}}
\put(7200,-2486){\makebox(0,0)[lb]{\smash{\SetFigFont{12}{14.4}{rm}{\color[rgb]{0,0,0}$\ket{{\bf d_n}}$}%
}}}
\put(7221,-536){\makebox(0,0)[lb]{\smash{\SetFigFont{12}{14.4}{rm}{\color[rgb]{0,0,0}$\ket{{\bf d_1}}$}%
}}}
\put(7203,-1211){\makebox(0,0)[lb]{\smash{\SetFigFont{12}{14.4}{rm}{\color[rgb]{0,0,0}$\ket{{\bf d_2}}$}%
}}}
\put(7224,-3136){\makebox(0,0)[lb]{\smash{\SetFigFont{12}{14.4}{rm}{\color[rgb]{0,0,0}$\ket{c_1 +\a^T{\bf d_1}}$}%
}}}
\put(7199,-3686){\makebox(0,0)[lb]{\smash{\SetFigFont{12}{14.4}{rm}{\color[rgb]{0,0,0}$\ket{c_2 + \a^T {\bf d_2}}$}%
}}}
\put(7199,-4961){\makebox(0,0)[lb]{\smash{\SetFigFont{12}{14.4}{rm}{\color[rgb]{0,0,0}$\ket{c_n + \a^T {\bf d_n}}$}%
}}}
\put(601,-4936){\makebox(0,0)[lb]{\smash{\SetFigFont{12}{14.4}{rm}{\color[rgb]{0,0,0}$\ket{c_n}$}%
}}}
\end{picture}

%% file: uniformC.pstex_t
\begin{picture}(0,0)%
\includegraphics{uniformC.pstex}%
\end{picture}%
\setlength{\unitlength}{3947sp}%
\begingroup\makeatletter\ifx\SetFigFont\undefined
\def\x#1#2#3#4#5#6#7\relax{\def\x{#1#2#3#4#5#6}}%
\expandafter\x\fmtname xxxxxx\relax \def\y{splain}%
\ifx\x\y   
\gdef\SetFigFont#1#2#3{%
  \ifnum #1<17\tiny\else \ifnum #1<20\small\else
  \ifnum #1<24\normalsize\else \ifnum #1<29\large\else
  \ifnum #1<34\Large\else \ifnum #1<41\LARGE\else
     \huge\fi\fi\fi\fi\fi\fi
  \csname #3\endcsname}%
\else
\gdef\SetFigFont#1#2#3{\begingroup
  \count@#1\relax \ifnum 25<\count@\count@25\fi
  \def\x{\endgroup\@setsize\SetFigFont{#2pt}}%
  \expandafter\x
    \csname \romannumeral\the\count@ pt\expandafter\endcsname
    \csname @\romannumeral\the\count@ pt\endcsname
  \csname #3\endcsname}%
\fi
\fi\endgroup
\begin{picture}(6775,7149)(226,-6673)
\put(1726,-1636){\makebox(0,0)[lb]{\smash{\SetFigFont{12}{14.4}{rm}{\color[rgb]{0,0,0}$F$}%
}}}
\put(1726,239){\makebox(0,0)[lb]{\smash{\SetFigFont{12}{14.4}{rm}{\color[rgb]{0,0,0}$F$}%
}}}
\put(1726,-361){\makebox(0,0)[lb]{\smash{\SetFigFont{12}{14.4}{rm}{\color[rgb]{0,0,0}$F$}%
}}}
\put(2401,-3436){\makebox(0,0)[lb]{\smash{\SetFigFont{12}{14.4}{rm}{\color[rgb]{0,0,0}$U$}%
}}}
\put(4351,-3436){\makebox(0,0)[lb]{\smash{\SetFigFont{12}{14.4}{rm}{\color[rgb]{0,0,0}$U$}%
}}}
\put(3376,-3436){\makebox(0,0)[lb]{\smash{\SetFigFont{12}{14.4}{rm}{\color[rgb]{0,0,0}$U$}%
}}}
\put(6226,-3436){\makebox(0,0)[lb]{\smash{\SetFigFont{12}{14.4}{rm}{\color[rgb]{0,0,0}$V$}%
}}}
\put(226,-286){\makebox(0,0)[lb]{\smash{\SetFigFont{12}{14.4}{rm}{\color[rgb]{0,0,0}$\ket{a_2}$}%
}}}
\put(226,314){\makebox(0,0)[lb]{\smash{\SetFigFont{12}{14.4}{rm}{\color[rgb]{0,0,0}$\ket{a_1}$}%
}}}
\put(226,-1547){\makebox(0,0)[lb]{\smash{\SetFigFont{12}{14.4}{rm}{\color[rgb]{0,0,0}$\ket{a_{n-1}}$}%
}}}
\put(226,-2594){\makebox(0,0)[lb]{\smash{\SetFigFont{12}{14.4}{rm}{\color[rgb]{0,0,0}$\ket{a_n}$}%
}}}
\put(226,-3344){\makebox(0,0)[lb]{\smash{\SetFigFont{12}{14.4}{rm}{\color[rgb]{0,0,0}$\ket{b}$}%
}}}
\put(5176,-3436){\makebox(0,0)[lb]{\smash{\SetFigFont{12}{14.4}{rm}{\color[rgb]{0,0,0}$\mathcal{I}$}%
}}}
\put(901,-5611){\makebox(0,0)[lb]{\smash{\SetFigFont{12}{14.4}{rm}{\color[rgb]{0,0,0}$\ket{\a}$}%
}}}
\put(976,-6436){\makebox(0,0)[lb]{\smash{\SetFigFont{12}{14.4}{rm}{\color[rgb]{0,0,0}$\ket{c}$}%
}}}
\put(6901,-5611){\makebox(0,0)[lb]{\smash{\SetFigFont{12}{14.4}{rm}{\color[rgb]{0,0,0}$\ket{\varphi}$}%
}}}
\put(6901,-6436){\makebox(0,0)[lb]{\smash{\SetFigFont{12}{14.4}{rm}{\color[rgb]{0,0,0}$\ket{a_n}$}%
}}}
\put(7001,-1611){\makebox(0,0)[lb]{\smash{\SetFigFont{14}{16.8}{rm}{\color[rgb]{0,0,0}$F\ket{a_{n-1}}$}%
}}}
\put(7001,-3411){\makebox(0,0)[lb]{\smash{\SetFigFont{14}{16.8}{rm}{\color[rgb]{0,0,0}$\ket{\varphi}$}%
}}}
\put(7001,-2661){\makebox(0,0)[lb]{\smash{\SetFigFont{14}{16.8}{rm}{\color[rgb]{0,0,0}$\ket{a_n}$}%
}}}
\put(7001,-336){\makebox(0,0)[lb]{\smash{\SetFigFont{14}{16.8}{rm}{\color[rgb]{0,0,0}$F\ket{a_2}$}%
}}}
\put(7001,264){\makebox(0,0)[lb]{\smash{\SetFigFont{14}{16.8}{rm}{\color[rgb]{0,0,0}$F\ket{a_1}$}%
}}}
\put(3751,-6136){\makebox(0,0)[lb]{\smash{\SetFigFont{29}{34.8}{rm}{\color[rgb]{0,0,0}$C$}%
}}}
\end{picture}

%% file: encode.pstex_t
\begin{picture}(0,0)%
\includegraphics{encode.pstex}%
\end{picture}%
\setlength{\unitlength}{3947sp}%
\begingroup\makeatletter\ifx\SetFigFont\undefined
\def\x#1#2#3#4#5#6#7\relax{\def\x{#1#2#3#4#5#6}}%
\expandafter\x\fmtname xxxxxx\relax \def\y{splain}%
\ifx\x\y   
\gdef\SetFigFont#1#2#3{%
  \ifnum #1<17\tiny\else \ifnum #1<20\small\else
  \ifnum #1<24\normalsize\else \ifnum #1<29\large\else
  \ifnum #1<34\Large\else \ifnum #1<41\LARGE\else
     \huge\fi\fi\fi\fi\fi\fi
  \csname #3\endcsname}%
\else
\gdef\SetFigFont#1#2#3{\begingroup
  \count@#1\relax \ifnum 25<\count@\count@25\fi
  \def\x{\endgroup\@setsize\SetFigFont{#2pt}}%
  \expandafter\x
    \csname \romannumeral\the\count@ pt\expandafter\endcsname
    \csname @\romannumeral\the\count@ pt\endcsname
  \csname #3\endcsname}%
\fi
\fi\endgroup
\begin{picture}(7350,3093)(226,-2623)
\put(5326,-361){\makebox(0,0)[lb]{\smash{\SetFigFont{12}{14.4}{rm}{\color[rgb]{0,0,0}$V_n$}%
}}}
\put(3076,-361){\makebox(0,0)[lb]{\smash{\SetFigFont{12}{14.4}{rm}{\color[rgb]{0,0,0}$U_n$}%
}}}
\put(226,-286){\makebox(0,0)[lb]{\smash{\SetFigFont{12}{14.4}{rm}{\color[rgb]{0,0,0}$\ket{\c}$}%
}}}
\put(226,314){\makebox(0,0)[lb]{\smash{\SetFigFont{12}{14.4}{rm}{\color[rgb]{0,0,0}$\ket{0}$}%
}}}
\put(1876,-61){\makebox(0,0)[lb]{\smash{\SetFigFont{12}{14.4}{rm}{\color[rgb]{0,0,0}$C$}%
}}}
\put(301,-1097){\makebox(0,0)[lb]{\smash{\SetFigFont{12}{14.4}{rm}{\color[rgb]{0,0,0}$\ket{\d}$}%
}}}
\put(301,-1769){\makebox(0,0)[lb]{\smash{\SetFigFont{12}{14.4}{rm}{\color[rgb]{0,0,0}$\ket{D}$}%
}}}
\put(376,-2444){\makebox(0,0)[lb]{\smash{\SetFigFont{12}{14.4}{rm}{\color[rgb]{0,0,0}$\ket{0^n}$}%
}}}
\put(7576,-361){\makebox(0,0)[lb]{\smash{\SetFigFont{14}{16.8}{rm}{\color[rgb]{0,0,0}$\ket{\varphi_{\c,\d}}$}%
}}}
\end{picture}